\documentclass[aps,prb,twocolumn,superscriptaddress,10pt,floatfix]{revtex4-2}

\usepackage{graphicx}
\usepackage{dcolumn}
\usepackage{bm}
\usepackage{hyperref}
\hypersetup{
    unicode=false,          
    pdftoolbar=false,        
    pdfmenubar=true,        
    pdffitwindow=false,     
    pdfstartview={FitH},    
    pdfkeywords={MBL} {AGP} {Mass-deformed SYK}, 
    pdfnewwindow=true,      
    colorlinks=true,       
    linkcolor=red,          
    citecolor=blue,        
    filecolor=magenta,      
    urlcolor=blue           
}

\usepackage{xcolor}
\usepackage{mathrsfs}
\usepackage{subfigure}
\usepackage{amsfonts}
\newcommand{\reffig}[1]{\mbox{Fig.~\ref{#1}}}
\newcommand{\T}{\({\mathcal T}\,\)}
\usepackage{physics}
\usepackage[page]{appendix}

\newcommand{\pcsadd}{Center for Theoretical Physics of Complex Systems, Institute for Basic Science (IBS), Daejeon - 34126, Korea}
\newcommand{\ustadd}{Basic Science Program, Korea University of Science and Technology (UST), Daejeon - 34113, Korea}

\newcommand{\mh}{\hat{\mathcal{H}}}
\newcommand{\hc}{\hat{\chi}}

\newcommand{\ma}{\mathcal{A}}
\newcommand{\malt}{\|\ma_\lambda\|^2}
\newcommand{\malkt}{\|\ma_\lambda(\kappa)\|^2}
\newcommand{\malzt}{\|\ma_\lambda(0)\|^2}
\newcommand{\tmalkt}{\|\tilde{\ma}_\lambda(\kappa)\|^2}

\newcommand{\tth}{\tau_{\mathrm{Th}}}

\begin{document}

\title{Delayed Thermalization in Mass-Deformed SYK}

\author{Dillip Kumar Nandy}
    \email{nandy@ibs.re.kr}
    \affiliation{\pcsadd}

\author{Tilen Čadež}
    \email{tilencadez@ibs.re.kr}
    \affiliation{\pcsadd}

\author{Barbara Dietz}
    \email{barbara@ibs.re.kr}
    \affiliation{\pcsadd}

\author{Alexei Andreanov}
    \email{aalexei@ibs.re.kr}
    \affiliation{\pcsadd}
    \affiliation{\ustadd}

\author{Dario Rosa}
    \email{dario\_rosa@ibs.re.kr}
    \affiliation{\pcsadd}
    \affiliation{\ustadd}

\date{\today}

\begin{abstract}
We study the thermalizing properties of the mass-deformed SYK model, in a regime of parameters where the eigenstates are ergodically extended over just portions of the full Fock space, as an all-to-all toy model of many-body localization.
Our numerical results strongly support the hypothesis that, although considerably delayed, thermalization is still present in this regime.
Our results add to recent studies indicating that many-body localization should be interpreted as a strict Fock-space localization.
\end{abstract}

\maketitle


\section{Introduction} 
\label{sec:intro}

The prospect of finding isolated quantum many-body systems capable to escape from thermalization, currently understood in the framework of the \emph{Eigenstate Thermalization Hypothesis} (ETH)~\cite{deutsch1991quantum, srednicki1994chaos, dalessio2018from}, has been one of the central aspects of research in condensed matter physics over the last two decades.
In typical thermalizing systems, any memory of an initial configuration quickly evolves into highly non-local degrees of freedom and is lost to large extent.
Identifying systems defying thermalization has become even more pressing in the last few years, with the recent advances in building quantum computers and other nano-devices.
Such systems, able to preserve information about their initial configurations, are of uttermost importance in order to implement concrete realizations of quantum memory.

An obvious way to evade thermalization is to consider integrable systems. 
However, they are generically very sensitive to perturbations and even small deviations, which are unavoidable in practical realizations, are enough to spoil integrability and restore thermalization.
It has been understood since the seminal work of Anderson, that disorder can provide a robust mechanism to avoid  thermalization~\cite{anderson1958absence}. 
Anderson localization has been studied extensively in subsequent years, leading to many numerical and exact results which are by now well-established and accepted~\cite{kramer1993review,evers2008anderson}.
For example, it has been proven mathematically that in any dimension and for large enough disorder localization occurs~\cite{hundertmark2008a} and that, in \(d = 1\), any infinitesimal uncorrelated disorder is enough to induce localization~\cite{thouless1972a}.
The same behavior is believed to hold in \(d = 2\) as well~\cite{anderson1980new}, although a rigorous proof is still missing, while in \(d = 3\) the presence of a finite critical disorder has been largely established~\cite{kramer1993review,evers2008anderson}. 
However the Anderson problem is, by definition, a single particle problem and, as such, is not describing a realistic situation with many particles and interactions that cannot be neglected.
Hence, it is natural to ask about the fate of the Anderson localization in presence of interactions.

To begin with, one has to define localization in a genuine many-body setup, since the single particle wave functions loose their significance and one cannot simply state that a certain wave function is localized in space~\cite{bardarson2012unbounded}.
This problem has initially been addressed through the notion of \emph{Fock-space localization}~\cite{altshuler1997quasi, gornyi2005interacting, basko2006metal}.
It is based on the observation that a generic disordered interacting many-body problem can be recast as an Anderson-like problem, but living on a highly connected graph defined in the Fock space.
There, the study of the possibility of having eigenstates which are localized in Fock space makes sense.
As a consequence, systems enjoying Fock-space localization, usually referred to as many-body localization (MBL), display peculiar features making them very different from systems obeying ETH.
Examples are the absence of transport~\cite{fleishman1980interaction}, area-law entangled eigenstates~\cite{serbyn2013local, bauer2013area}, 
a logarithmic growth of entanglement entropy after a quantum quench~\cite{znidaric2008many-body, bardarson2012unbounded}, Poissonian spectral correlations~\cite{oganesyan2007localization, pal2010many-body}, the emergence of an extensive set of \emph{local integrals of motion} (LIOMs)~\cite{serbyn2013local, ros2015integrals, huse2014phenomenology} and others.
It was later argued that strict Fock-space localization was not necessary to display the phenomenological features of MBL listed above,~\cite{bauer2013area}, and that even in presence of extended eigenstates in Fock space, but with an extension much smaller than the dimension of the Hilbert space, the area-law entanglement entropy may emerge.
This point of view was later confirmed numerically by the authors of Ref.~\onlinecite{luitz2015many}, who investigated the paradigmatic example of the disordered Heisenberg model in one dimension~\cite{oganesyan2007localization, pal2010many-body}.
According to their results, eigenstates in the MBL phase have non-trivial fractal dimension in Fock space and, in particular, they are not localized.

However the absence of simple, solvable models, makes the understanding of MBL extremely challenging.
Many results are either derived relying on some approximations or obtained numerically for relatively small system sizes, accessible to classical computers, which might not be enough to reach the thermodynamic limit~\cite{suntas2020quantum, suntas2020ergodicity, abanin2021distinguishing, sels2021dynamical, morningstar2022avalanches}.

A notable exception is represented by the so-called \emph{mass-deformed SYK model}~\cite{garcia-garcia2018chaotic, nosaka2018the}, which is a fully-connected disordered interacting model modified to include a random mass-term. 
This model, depending on the strength of the mass deformation, shows a transition from  ergodic to  localized in Fock space which can be studied analytically in the large-\(N\) limit~\cite{monteiro2021minimal, monteiro2021quantum}.
In addition, for intermediate values of the mass deformation, eigenstates are extended (ergodically) in Fock space but their support is restricted to domains which are much smaller than the full Hilbert space dimension.
We will refer to this regime as the \emph{cluster regime}.
Hence, this model represents an excellent toy model of MBL, including the clustering property emphasized in Refs.~\onlinecite{bauer2013area} and~\onlinecite{luitz2015many}, despite the fact of the model not having any spatial extension.

In this paper we re-analyze numerically the mass-deformed SYK model, with a particular focus on understanding to which extent its ergodic properties are absent or present in the cluster regime.
The numerical analysis is performed by means of the Adiabatic Gauge Potential (AGP), recently introduced in Ref.~\onlinecite{pandey2020adiabatic} as a very sensitive probe of quantum chaos, based on the response of eigenvectors under small deformations of the model.
In agreement with the analysis of Ref.~\onlinecite{sels2021dynamical} we quantitatively find that the cluster regime can be characterized as a region where the scaling of AGP with system size is \emph{faster} than the scaling predicted by ETH.
In addition, we investigate the ergodic properties of this regime by studying the Spectral Form Factor (SFF) and the associated notion of Thouless time.
Our findings show that the Thouless time has a scaling with system size which, for large enough systems, is comparable with that in the ergodic regime.
Thus, our results suggest that, at least for this particular model, the cluster regime has to be considered as delayed, but still thermalizing regime.
On the other hand, to observe a genuine violation of thermalization, one has to consider the regime of Fock-space localization.

The paper is organized as follows.
In Sec.~\ref{sec:model} we recall the main features of the mass-deformed SYK model, following the results of Refs.~\onlinecite{monteiro2021minimal} and \onlinecite{monteiro2021quantum}.
In Sec.~\ref{sec:AGP} we present our numerical results based on the evaluation of AGP.
In Sec.~\ref{sec:SFF} we use the SFF to show that the intermediate regime, in which eigenstates are extended over clusters of dimension scaling with \(N\), is thermalizing.
In Sec.~\ref{sec:conclusions} we conclude our findings.
Finally, in the Appendices we present a more detailed analysis of the random matrix theory properties of the model under investigation (App.~\ref{app:spectral}) as well as a more detailed discussion of the SFF definition (App.~\ref{app:sff}).

\section{The model} 
\label{sec:model}

The model under consideration has been dubbed in  literature as ``\emph{mass-deformed}'' SYK,~\cite{garcia-garcia2018chaotic, kim2021comment, garcia-garcia2021reply, nosaka2018the, monteiro2021minimal}.
As the name suggests, it can be thought of as a deformation of the celebrated SYK model~\cite{sachdev1993spin, maldacena2016remarks, polchinski2016the}, by adding a quadratic random mass term.
It is realized in terms of \(N\) Majorana fermions, \(\hc_i, \ i = 1, \dots, N\), i.e. quantum mechanical operators satisfying Clifford algebra relations \(\left\{ \hc_i,\, \hc_j\right\} = \delta_{ij}\).
The model is given by the following Hamiltonian
\begin{align}
    \label{eq:mass_deformed_hamiltonian}
    & \mh \equiv \frac{2}{\sqrt{N}}\mh_4 + \kappa \, \mh_2, \notag \\
    & \mh_4 = -\sum_{i<j<k<l} J_{ijkl} \hc_i \hc_j \hc_k \hc_l, \; \mh_2 = i \sum_{i<j} J_{ij} \hc_i \hc_j \, ,
\end{align}
where the coupling constants \(J_{ijkl}\) and \(J_{ij}\) are Gaussian distributed with vanishing mean values and variances given by \(\frac{6}{N^3}\) and \(\frac{1}{N}\), respectively.
Finally, \(\kappa\) is the mass-deformation strength parameter, which controls the strength of the quadratic mass deformation compared to the quartic interaction term.

The model above has been studied in Refs.~\onlinecite{monteiro2021minimal} and \onlinecite{monteiro2021quantum} as an analytically tractable model of Fock-space localization.
As such, we are going to use it as a platform to benchmark various probes of the ETH/MBL transition used in literature at finite values of \(N\) versus their large-\(N\) predictions in the thermodynamic limit.
More in details, Ref.~\onlinecite{monteiro2021minimal} identified \(4\) major regimes as a function of \(\kappa\) (see also Ref.~\onlinecite{micklitz2019nonergodic}):
\begin{itemize}
    \item[---] Regime \(I\): \(\kappa < \sqrt{\frac{(N - 2) (N - 3)}{2N^3}}\).
    Eigenstates are ergodically extended over the full Fock space and ETH holds for all eigenstates.
    \item[---] Regimes \(II\) and \(III\) -- \emph{cluster regime}: \(\sqrt{\frac{(N - 2) (N - 3)}{2N^3}} < \kappa < \frac{Z}{2\sqrt{2 \rho}} W \left( 2Z \sqrt{\pi} \right)\). 
    Eigenstates are ergodically extended over energy shells whose dimension scales \emph{exponentially} in \(N\).
    This is the most interesting regime: a given eigenstate, in the Fock space, is extended to cover just the nearest neighbors. 
    Such a configuration, in standard Anderson problems in finite dimensions, would give rise to localization. 
    However, since the connectivity of the mass-deformed model in the Fock space scales with \(N\), these states still have an extensive support in Fock space.
    The distinction between the two regimes, as discussed in Ref.~\onlinecite{monteiro2021minimal} is mostly quantitative rather than qualitative.
    The boundary between the two regions is located at \(\kappa = \sqrt{\frac{6}{ N^4} \binom{N}{4}} \).
    \item[---] Regime \(IV\) -- Fock-space localization: \(\kappa >\frac{Z}{2\sqrt{2 \rho}} W \left( 2Z \sqrt{\pi} \right)\).
    Eigenstates are localized in Fock space.
\end{itemize}
We define \(Z \equiv \binom{N/2}{4}\) and \(\rho \equiv \binom{N}{4}\), and \(W\) is the Lambert \(W\) function.

\section{Numerical results: Adiabatic Gauge Potential} 
\label{sec:AGP}

The Adiabatic Gauge Potential has been proposed as a very sensitive probe of quantum chaos~\cite{pandey2020adiabatic}.
It measures the sensitivity of eigenstates to small perturbations of the Hamiltonian.
In chaotic models, eigenstates are expected to be very sensitive to deformations, in distinction to integrable systems where such a sensitivity is supposed to be much smaller~\cite{kolodrubetz2017geometry}.
More quantitatively, let us consider a given Hamiltonian \(\mh\) with eigenvectors \(\ket{n}\).
We add a small deformation to \(\mh\), \( \mh \to \mh(\lambda) = \mh + \lambda \hat{O} \), where \(\hat O\) is a generic operator which does not commute with \(\mh\)  and \(\lambda\) is a parameter.
Then the AGP, denoted by \(\malt\) is defined as follows:
\begin{align}
    \label{eq:AGP_def}
    \malt &\equiv \frac{1}{D} \sum_{m \neq n} \abs{\bra{n}\hat{\ma}_\lambda (\mu)\ket{m}}^2 , \notag \\
    \bra{n}\hat{\ma}_\lambda (\mu)\ket{m} &\equiv - i \frac{\omega_{nm}}{\omega_{mn}^2 + \mu^2} \, \bra{n} \hat O \ket{m} ,
\end{align}
where \(\omega_{mn}\) denotes the energy difference, \(\omega_{mn} \equiv E_m - E_n \), \(\mu\) is a regularizing cutoff at \(E_m \sim E_n\) and \(D\) is the Hilbert space dimension.

The authors of Ref.~\onlinecite{pandey2020adiabatic} studied extensively the scaling properties of \(\malt\) for several examples of integrable and chaotic systems.
Their findings indicate that, in agreement with ETH, \(\malt\) scales \emph{exponentially} with the system size \(N\), when the system is thermalizing.
On the other hand, the scaling with the system size is much milder for integrable systems, including the extreme case of free particles for which \(\malt\) does not scale at all with the system size.

In the case at hand, we studied the behavior of \(\malkt\), as a function of \(\kappa\), for the following extensive operator 
\begin{gather}
    \label{eq:operator_AGP_extensive_def}
    \hat{O} \equiv \sum_{i = 1}^{N - 1} \hc_i \hc_{i + 1} .
\end{gather}
In addition, we have checked that our results are robust against the choice of the operator by employing other operators, including non-extensive ones, and found no qualitative differences.

Obviously, for \(\kappa = 0\) we simply recover the usual SYK model for which ETH holds~\cite{hunter-jones2018on}. 
Hence, to better investigate deviations away from the ETH behavior, we define the following modification of the AGP
\begin{equation}
    \label{eq:AGP_renormalized_def}
    \tmalkt \equiv \frac{\malkt}{\malzt}.
\end{equation}
which we refer to simply as AGP in the following.
From its definition, \(\tmalkt\) is expected to be \emph{\(N\)-independent} in the ETH region, \emph{i.e.} for small values of \(\kappa\).
On the other hand, for very large values of \(\kappa\), where Fock-space localization sets in, \(\malkt\) is expected to be largely independent on \(N\), so that \(\tmalkt\) is expected to scales down exponentially with system size.

\begin{figure}[t!]
\begin{center}
    \includegraphics[width=\columnwidth]{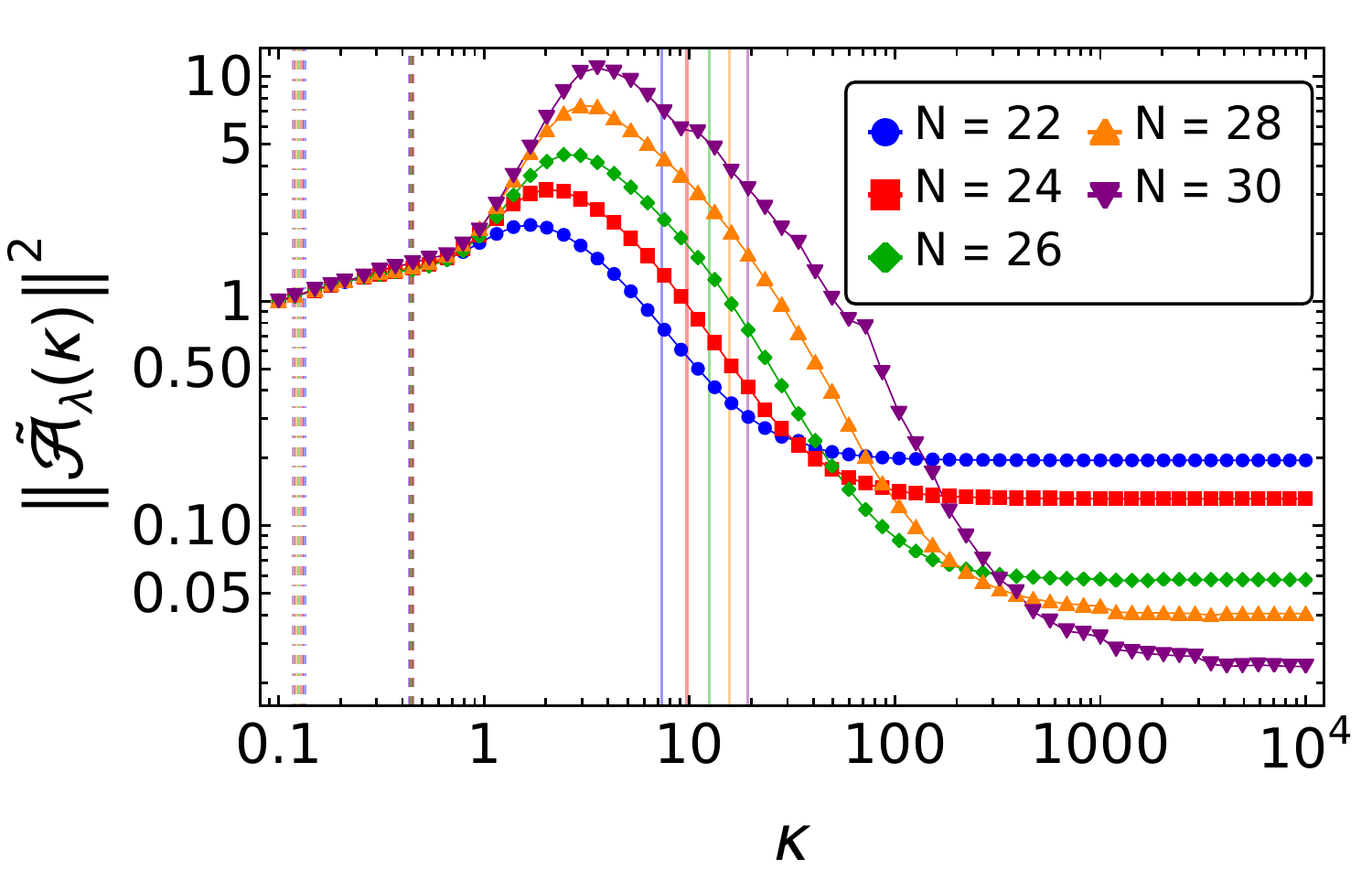}\\
    \caption{
        The behavior of AGP, as a function of the mass deformation strength \(\kappa\). 
        Results are obtained by considering infinite temperature eigenstates, \emph{i.e.} eigenstates having energies \(-0.05 < E < 0.05\).
        Vertical lines show regime boundaries (colors reflect the \(N\) dependence of such boundaries):
            solid lines refer to boundaries between regime \(III\) and regime \(IV\), 
            dashed lines to boundaries between regime \(II\) and regime \(III\),  and 
            dot-dashed lines to boundaries between regime \(I\) and \(II\).
    }  
    \label{fig:AGP_plot_dillip_data}
\end{center}
\end{figure}

Based on these premises, we have computed \(\tmalkt\), for system sizes ranging from \(N = 22\) to \(N = 30\).
Results of our numerical analysis are summarized in Fig.~\ref{fig:AGP_plot_dillip_data}.
We see that both Regime \(I\) and Regime \(II\) follow quite closely the prediction of ETH, with the AGP being \(N\)-independent and the curves collapsing on top of each other.
Similarly, for large values of disorder, the AGP is \(\kappa\)-independent and inversely proportional to \(N\), as expected for a free theory.
Interestingly, we observe that the value of \(\kappa\) at which Fock-space localization becomes manifest is systematically larger than the large-\(N\) value predicted in Ref.~\onlinecite{monteiro2021minimal}.
Such a behavior is actually in agreement with the results of Ref.~\onlinecite{pandey2020adiabatic}, where it was shown that AGP is an extremely sensitive probe for quantum chaos, compared to other standard spectral diagnostics.

Much more interesting is the behavior of the AGP for larger but still intermediate values of \(\kappa\), corresponding to the late cluster regime in Regime \(III\).
In this regime, we find that \(\tmalkt\) scales with \(N\) \emph{faster} than predicted by ETH, as is clear from the fact that curves for different values of \(N\) do not collapse on top of each other.
This kind of behavior has been already reported in Ref.~\onlinecite{sels2021dynamical}, where it was attributed to a glassy-like dynamics and \emph{not} to a genuine MBL phase.
Interestingly, the numerical distinction between Regime \(II\) and Regime \(III\) was very challenging to observe in Ref.~\onlinecite{monteiro2021minimal}.
On the other hand, AGP seems to be very sensitive to this change of regime.

All in all, our results are in excellent agreement with Refs.~\onlinecite{sels2021dynamical} and \onlinecite{leblond2021universality}, while they only partially agree with the analytical predictions of Ref.~\onlinecite{monteiro2021minimal}. 
From finite size analysis it is possible to clearly distinguish just \emph{three} regimes, instead of the four predicted in the thermodynamic limit: an ergodic regime, a regime with a scaling larger than ETH (cluster regime) and finally the Fock space localized regime.

\section{Numerical results: Spectral Form Factor} 
\label{sec:SFF}

\begin{figure*}
    \centering
    \subfigure[]{
        \label{fig:SFF_ergodic}
        \includegraphics[width=\columnwidth]{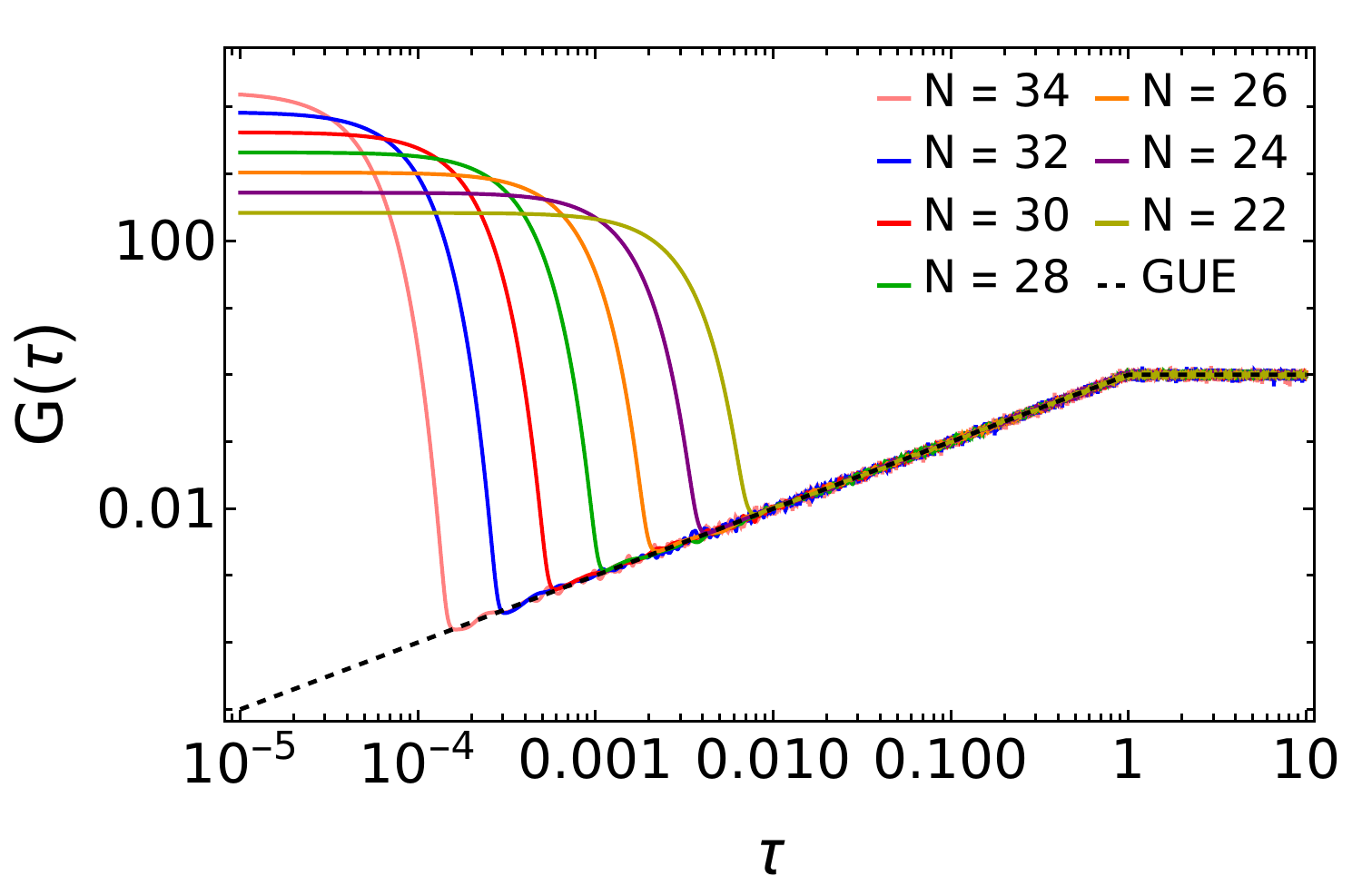}
    }
    \subfigure[]{
        \label{fig:SFF_glassy}
        \includegraphics[width=\columnwidth]{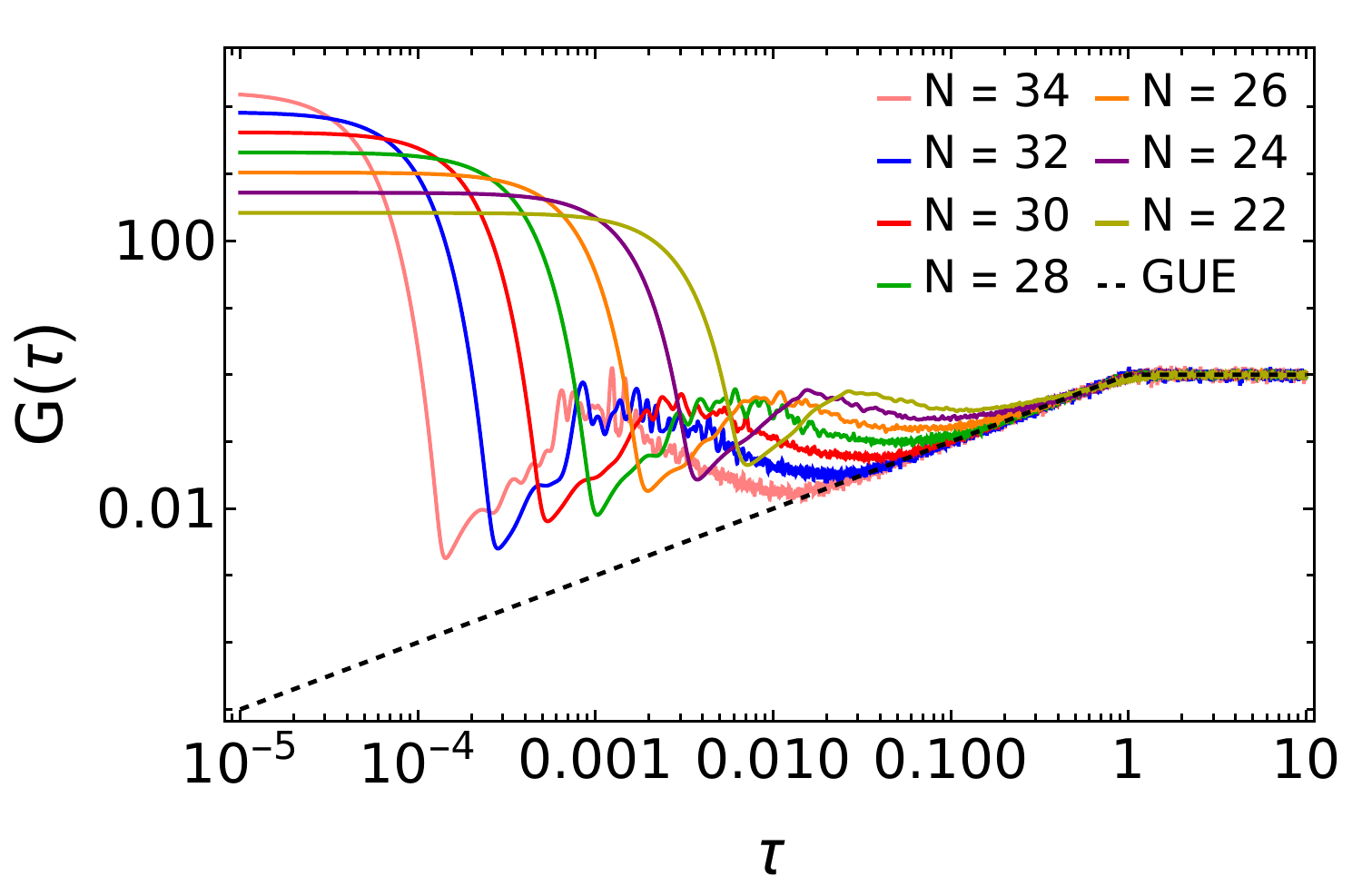}
    }
    \subfigure[]{
        \label{fig:SFF_localized}
        \includegraphics[width=\columnwidth]{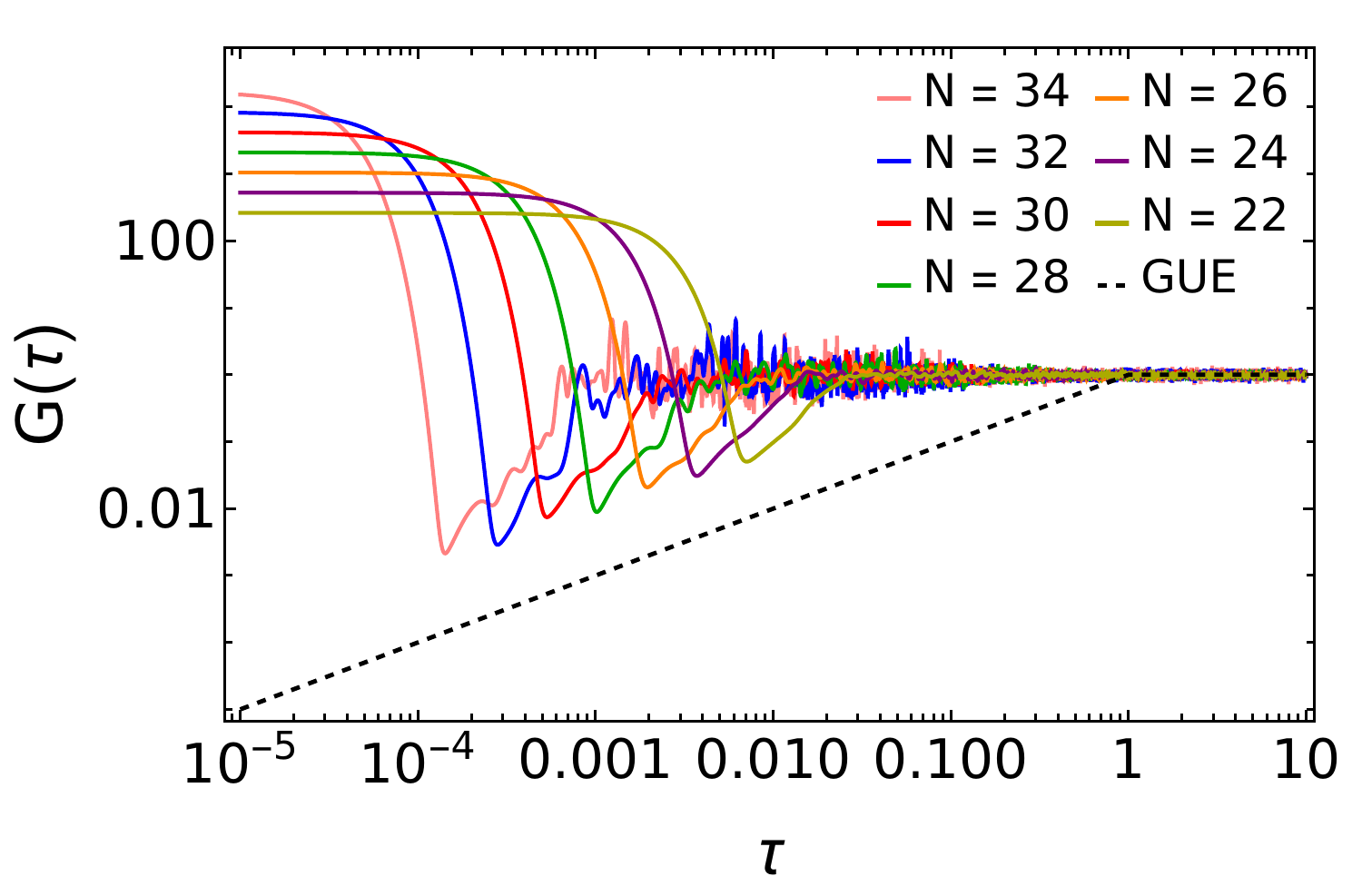}
    }
    \subfigure[]{
        \label{fig:SFF_ratios}
        \includegraphics[width=\columnwidth]{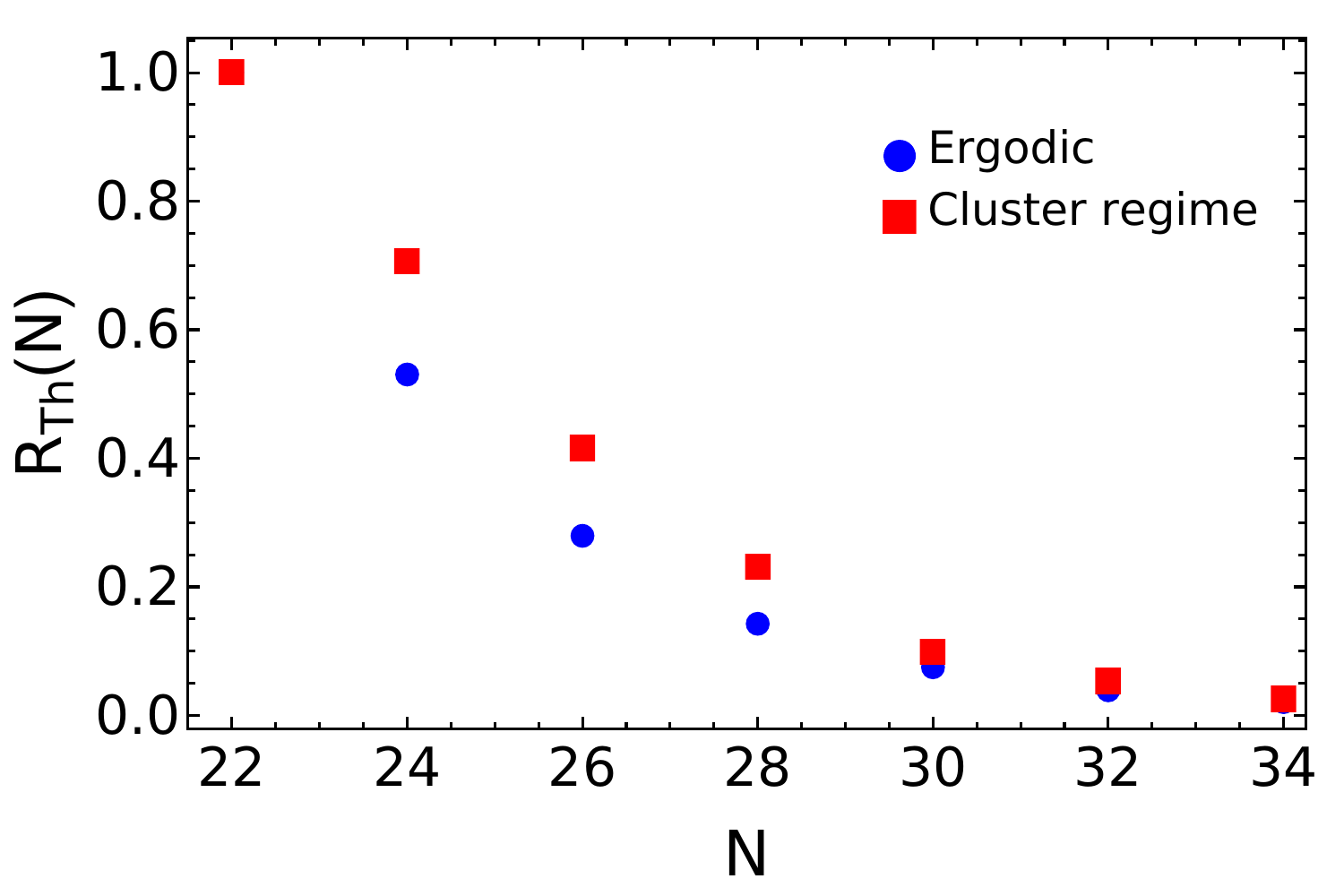}
    }
    \caption{
        Spectral form factor in: (a) ergodic (I); (b) cluster (III); and (c) Fock-space localized (IV) regimes.
        While in the ergodic regime the SFF follows clearly the standard RMT behaviour, in the cluster regime (regime III), the onset of the RMT behaviour is considerably delayed.
        The Fock-space localized regime does not show any RMT behaviour.
        (d) Dependence of \(R_{\mathrm{Th}} (N)\), Eq.~\eqref{eq:Thouless_ratio}, vs system size \(N\).
    }
\end{figure*}

Having found that, in the late cluster regime, AGP has a faster scaling with \(N\), we now probe in more detail the thermalizing properties of this regime.
Thermalization timescales are often associated with the so-called \emph{Thouless time}, \(\tth\),  which is by definition the time after which quantum evolution becomes universal and is described by Random Matrix Theory (RMT).
The Thouless time has been extensively studied recently via the SFF~\cite{cotler2017black, garcia-garcia2018universality, gharibyan2018onset}.
In particular, Ref.~\onlinecite{nosaka2018the} investigated some aspects of the Thouless time for the mass deformed SYK.
However, there the focus was mainly on the low temperature region of the SYK spectrum and on the possible hints of scrambling dynamics in the SFF, while an in depth analysis of the thermalizing properties in regime \(III\) was not performed. This will be done in this section.

By definition, the SFF is the Fourier transform of the density-density correlation function.
For computational purposes, it is more convenient to consider a Gaussian filtered version of it~\cite{gharibyan2018onset}.
Concretely, we adopt the definition of Ref.~\onlinecite{suntas2020quantum}
\begin{gather}
    \label{eq:SFF_def}
    G(\tau) \equiv \frac{1}{Z} \left\langle \abs{\sum_{n = 1}^D \rho(\epsilon_n) e^{- i 2 \pi \epsilon_n \tau}}^2 \right\rangle,
\end{gather}
where \(Z\) is a normalizing factor to ensure \(G(\tau \gg 1) \approx 1\), the \(\epsilon_n\) are the unfolded energy levels and \(\rho(\epsilon_n)\) is the Gaussian filtering function; see App.~\ref{app:sff} for more details. The time is measured in units of the Heisenberg time.

Given these preliminaries, we have computed the SFF, for increasing system sizes, for three values of \(\kappa\), \emph{i.e.} \(\kappa = 0.1, \, 2.09, \, 100 \). They have been chosen to be well-inside the ergodic regime, regime \( III \) and regime \(IV\), respectively~\footnote{since the SFF is a spectral observable, 
\emph{i.e.} it is based on eigenvalues, the relevant separation between regime \(III\) and regime \(IV\) is the one described by analytical predictions (vertical solid lines in Fig.~\ref{fig:AGP_plot_dillip_data}).}.
Our results are presented in Fig.~\ref{fig:SFF_ergodic} -- \ref{fig:SFF_localized}.
Results for \(k = 0.1\) and \(100\) are in agreement with the expectations: for \(k = 0.1\) the SFF shows a robust ramp behavior, in perfect agreement with the GUE prediction and with the Thouless time decreasing with increasing system size.
Similarly, when \(\kappa = 100\) the SFF does not show any agreement with RMT predictions, with \(\tth\) becoming identical to the Heisenberg time.

Once again, the interesting results appear for \(\kappa = 2.09\).
We see that at early times the SFF shows the usual initial decay, followed by an intermediate regime which \emph{is not} controlled by RMT-like behavior.
Interestingly, this early time behavior appears to be in very good agreement with the analogue behavior exhibited by the SFF at large disorder, \(\kappa = 100\), for which the full model is essentially controlled by the quadratic SYK\(_2\) Hamiltonian.
In particular, a faster than RMT ramp is clearly visible.
Here we stress that, for the pure SYK\(_2\) model, the presence of this fast ramp (exponential in time) can be computed analytically in the large-\(N\) limit~\cite{winer2020exponential, liao2020many, liao2022universal}.
Our analysis shows that the exponential ramp dominates the behavior of the SFF also in a regime of intermediate disorder.
However, at later times, but still earlier than the Heisenberg time, the SFF approaches the RMT predictions, showing a ramp in perfect agreement with GUE behavior.
The SFF decay --- from the exponential ramp to the RMT ramp --- looks reminiscent of the transition experienced by the SFF for a bunch of single particle chaotic models when coupled by nearest neighbor interactions, as described in Ref.~\onlinecite{chan2018spectral}.
Overall, we see this as a consequence of the interplay between the \(\mathrm{SYK}_2\) Hamiltonian and the interacting quartic term, so that the Thouless time gets inflated by, approximately, two orders of magnitude (in units of Heisenberg time).

To better quantify the ergodic properties of the SFF in this regime, and to compare them with the analogue properties of the SFF in the manifestly ergodic regime, we measured the scaling of the Thouless time with \(N\), \emph{in units of the Thouless time measured at} \(N = 22\)~\footnote{The case \(N = 22\) seems to be smallest value for which a scaling behavior is observed~\cite{monteiro2021minimal}.}.
In other words, we define the following quantity
\begin{gather}
    \label{eq:Thouless_ratio}
    R_{\mathrm{Th}} (N) \equiv \frac{\tth (N)}{ \tth (22)},
\end{gather}
and our results are reported in Fig.~\ref{fig:SFF_ratios}.
As we see, for small system sizes the scaling of \(R_{\mathrm{Th}}\) with \(N\) is faster in the ergodic regime.
However the situation starts to change for \(N \geq 30\) where the scaling in the two regimes becomes comparable.
This result strongly supports those of Ref.~\onlinecite{sels2021dynamical}, according to which the regime where AGP scaling is faster than ergodic is a \emph{delayed but still thermalizing} regime.

\section{Conclusions and outlook} 
\label{sec:conclusions}

In this paper, we have studied the thermalizing properties of mass-deformed SYK, in a regime of intermediate disorder, relying on the analytical results of Ref.~\onlinecite{monteiro2021minimal}.
In quantitative terms, this regime (dubbed as cluster regime in this paper), is characterized by eigenstates which are extended but cover only a fraction of the total Hilbert space.
It has been argued, in Ref.~\onlinecite{bauer2013area} and further confirmed by several subsequent numerical studies (see for example Ref.~\onlinecite{luitz2015many}), that a proper MBL regime \emph{is not} a regime in which eigenstates are localized in Fock space -- the so-called Fock-space localization discussed in seminal MBL papers~\cite{altshuler1997quasi, gornyi2005interacting, basko2006metal} -- but it is a regime in which eigenstates are extended, but not ergodically spread over the full Hilbert space.
Hence, the cluster regime can be viewed as the all-to-all version of the MBL regime (although, in mass deformed SYK, eigenstates are ergodically extended over their support \cite{monteiro2021quantum}), thus making it particularly suitable for numerical analysis, to be contrasted with the analytical results presented in Ref.~\onlinecite{monteiro2021minimal}.

In agreement with Ref.~\onlinecite{sels2021dynamical}, we found that this regime of intermediate disorder has a very clear characterization in terms of the AGP norm, as defined in Eq.~\eqref{eq:AGP_renormalized_def}: it is the regime where the AGP shows a scaling with system size which is faster than ETH, dubbed as \emph{maximally chaotic} in Ref.~\onlinecite{sels2021dynamical}.

To better characterize the dynamical features of this regime, we have computed the SFF and have contrasted it with the SFFs computed in the manifestly ergodic and in the Fock-space localized regimes.
The SFF is particularly suitable for such an analysis, since it not only distinguishes chaotic spectra from non-chaotic ones, but it also provides an estimate of the timescale, the Thouless time \(\tth\), at which universality arises and can be detected.
Our analysis shows that in this intermediate regime \(\tth\) is delayed by roughly two orders of magnitude as compared to \(\tth\) computed in the manifestly ergodic regime.
However, although delayed it is still clearly smaller than the Heisenberg time.
In addition, we have studied the scaling of \(\tth\) with system size, as compared to the analogous scaling in the  manifestly ergodic regime.
While at smaller sizes the two scalings are different, these discrepancies become negligible for the largest sizes we could probe, \emph{i.e.} for \(N = 30, \, 32,\, 34\).

Overall, our results strongly suggest that the late cluster regime should be considered as a regime of delayed thermalization and, although such a delay is quantitatively large, so that it could be easily missed in time evolution analysis, the system still appears completely thermalizing.
To find a genuinely non-thermalizing regime, \emph{i.e.} a regime in which the Thouless time does coincide with the Heisenberg time, one has to go to Regime \(IV\), which is the regime of genuine Fock-space localization.

Our results, being in agreement with the analysis of Refs.~\onlinecite{sels2021dynamical} and \onlinecite{leblond2021universality} (and taking into account the analytical predictions of Ref.~\onlinecite{monteiro2021minimal}) add to other recent results questioning the usual idea according to which the MBL phase, in disordered models, should not be understood as Fock-space localization and that in the MBL phase eigenstates are actually extended and the  extension scales with system size \(N\).
In particular, the thermalizing properties of this regime seem to become more prominent with increasing system size.
We would also like to point out the possibility that the existence of non Fock-space localized MBL might depend on subtle model details, e.g. type of disorder and of interaction, as suggested for example in Ref.~\onlinecite{sierant2021challenges}.
On the other hand, we believe that these results clearly show that mass-deformed SYK does capture all the interesting features which are at the core of the recent debates on MBL.

Of course, our numerical results, being based on full exact diagonalization are limited to systems of finite size and cannot be taken as definitive results concerning the fate of MBL in the thermodynamic limit.
Along this line, it would be extremely important to fully understand the origin of the discrepancy between the analytical predictions of Ref.~\onlinecite{monteiro2021minimal}, predicting \emph{four} distinguished regimes, and the numerical analysis via AGP where just three regimes can be distinguished.
Perhpas, it could be possible to use the solvability of the mass-deformed SYK to investigate the behavior of AGP and SFF in the thermodynamic limit in the cluster regime.

Very recently a toy model has been presented showing a regime with eigenstates being non-ergodically extended -- unlike the mass-deformed SYK model -- and Poissonian statistics \cite{tang2021non-ergodic}.
This could be a good avatar to describe genuine models of MBL without Fock-space localization.
It will be interesting to characterize this regime through the lens of AGP.

We hope to come back to these points in the near future.

\section*{Acknowledgments} 
\label{sec:acknowledgments}

We acknowledge support by the Institute for Basic Science in Korea (IBS-R024-D1). 
We thank Masaki Tezuka for kindly explaining us the results of Refs.~\onlinecite{monteiro2021minimal} and \onlinecite{monteiro2021quantum}. 
We thank Boris Altshuler, David J. Luitz, Anatoli Polkovnikov and Lev Vidmar for interesting discussions and comments.

\appendix

\section{Spectral Properties}
\label{app:spectral}

We investigated fluctuation properties in the eigenvalue spectra for the three realizations \(\kappa =0.2,2.09,100\) with \(N=34\) considered in~\reffig{fig:SFF_ergodic}, \reffig{fig:SFF_glassy} and~\reffig{fig:SFF_localized}, referred to as cases A, B, and C in the following. 
They are compared with RMT results for random matrices from the GUE. 
For this we removed system-specific properties by unfolding the eigenvalues \(E_j,\, E_1\leq E_2\dots \leq E_D\), independently for each disorder realization, to mean spacing unity, \(\epsilon_j=\bar N(E_j)\). 
The mean integrated spectral density \(\bar N(E_j)\) is determined by fitting a polynomial of order \(12\) to the integrated spectral density.

We analyzed short-range correlations in the eigenvalue spectra in terms of the nearest-neighbor spacing distribution \(P(s)\) of adjacent spacings \(s_j=\epsilon_{j+1}-\epsilon_j\) and its cumulant \(I(s)=\int_0^sds^\prime P(s^\prime)\),
which has the advantage that it does not depend on the binning size of the histograms yielding \(P(s)\).
Another measure for short-range correlations is the distribution of the ratios~\cite{oganesyan2007localization,Atas2013} of consecutive spacings between \(l\)th nearest neighbors, 
\(r_j=\frac{\epsilon_{j+l}-\epsilon_{j}}{\epsilon_{j+l-1}-\epsilon_{j-1}}\), where we chose \(l=1,\dots ,10\) and the distribution of \(\tilde r=\min\{r_j,\frac{1}{r_j}\}\).
Ratios are dimensionless so the non-unfolded eigenvalues can be used~\cite{oganesyan2007localization,Atas2013,Atas2013a}.
Furthermore, we considered the variance \(\Sigma^2(L)=\left\langle\left( N(L)-\langle N(L)\rangle\right)^2\right\rangle\) of the number of unfolded eigenvalues \(N(L)\) in an interval of length \(L\), 
and the rigidity \(\Delta_3(L)=\left\langle\min_{a,b}\int_{e-L/2}^{e+L/2}de\left[N(e)-a-be\right]^2\right\rangle\) as measures for long-range correlations. 
Here, \(\langle\cdot\rangle\) denotes the average over an ensemble of random matrices or of 100 eigenvalue spectra, each containing \(65536\) levels. 
For the latter we also performed spectral averages. 
Furthermore, we analyzed the power spectrum which is given in terms of the Fourier transform of the deviation of the \(q\)th nearest-neighbor spacing from its mean value \(q, \delta_q=\epsilon_{q+1}-\epsilon_1-q\), from \(q\) to \(\tau\)
\begin{gather}
    \label{eq:PowerS}
    s(\tau) = \left\langle\left\vert\frac{1}{\sqrt{n}}\sum_{q=0}^{n-1} \delta_q\exp\left(-2\pi i\tau q\right)\right\vert^2\right\rangle
\end{gather}
for a sequence of \(n\) levels, where \(0\leq\tau\leq 1\).
It exhibits for \(\tau\ll 1\) a power law dependence \(\langle s(\tau)\rangle\propto \tau^{-\alpha}\)~\cite{Relano2002,Faleiro2004}, 
where for regular systems \(\alpha =2\) and for chaotic ones \(\alpha =1\) independently of whether \T invariance is preserved or not~\cite{Gomez2005,Salasnich2005,Santhanam2005,Relano2008,Faleiro2006,Mur2015}.

\begin{figure}[!th]
    \includegraphics[width=\columnwidth]{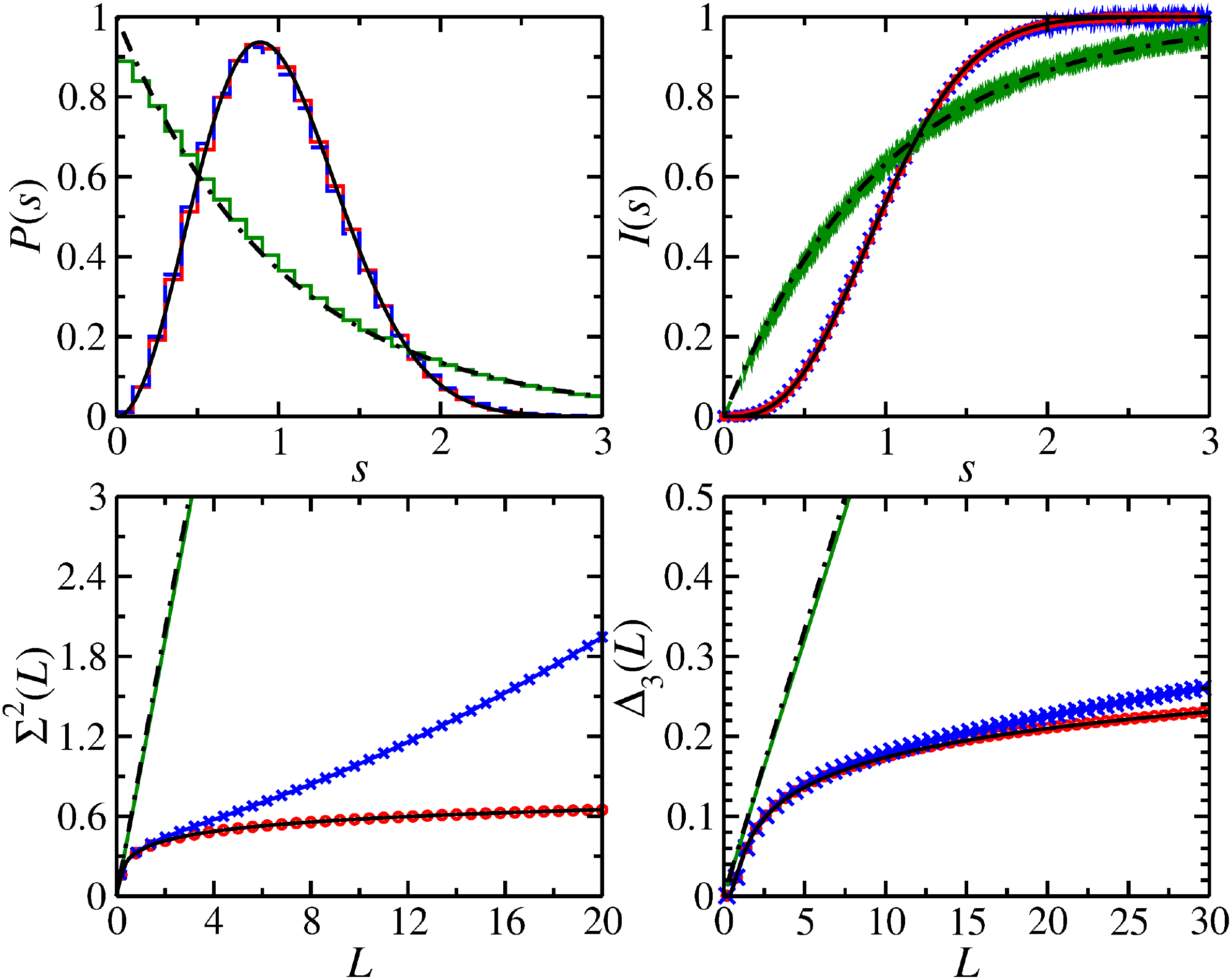}
    \caption{
        Comparison of the spectral properties of cases A (red histogram and circles), B (blue histogram and crosses) and C (green histogram and lines) the results for Poissonian random numbers from the GUE (black full lines) and Poissonian random numbers (black dash-dotted lines), respectively.
    }
    \label{fig:spectral}
\end{figure}

\begin{figure}[!th]
    \includegraphics[width=\columnwidth]{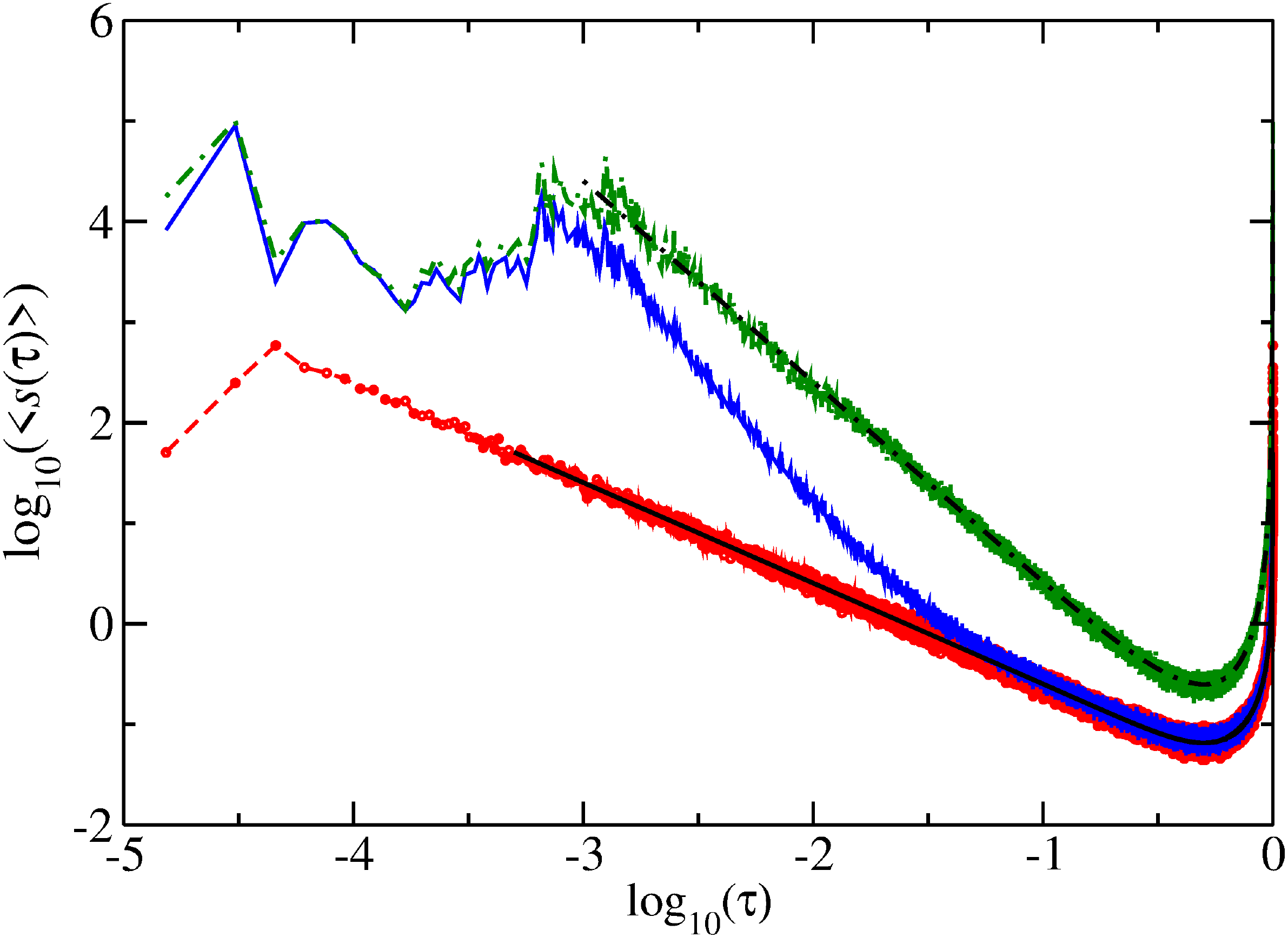}
    \caption{
        Same as~~\reffig{fig:spectral} for the power spectrum.
    }
    \label{fig:power}
\end{figure}

\begin{figure}[!th]
    \includegraphics[width=\columnwidth]{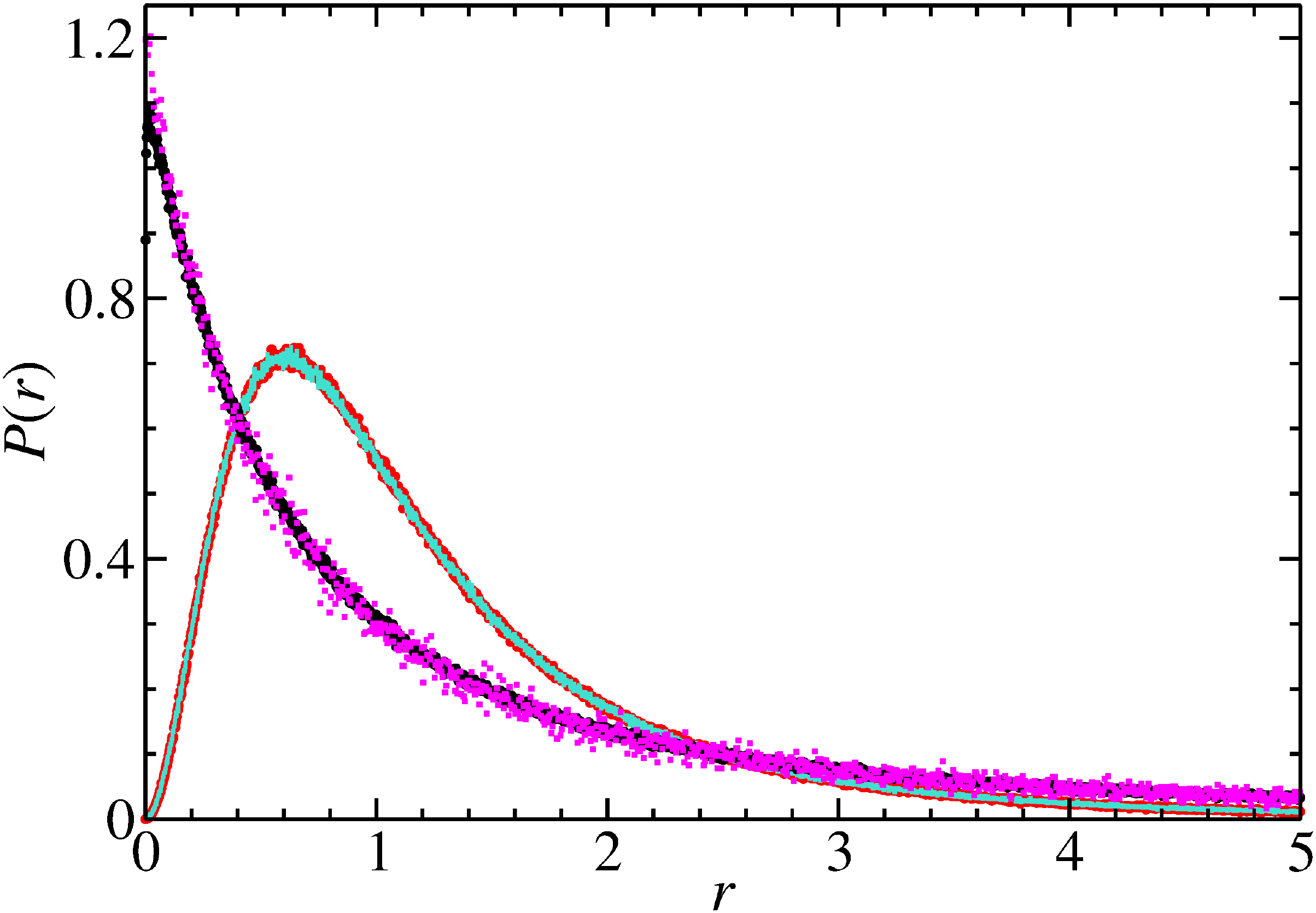}
    \includegraphics[width=\columnwidth]{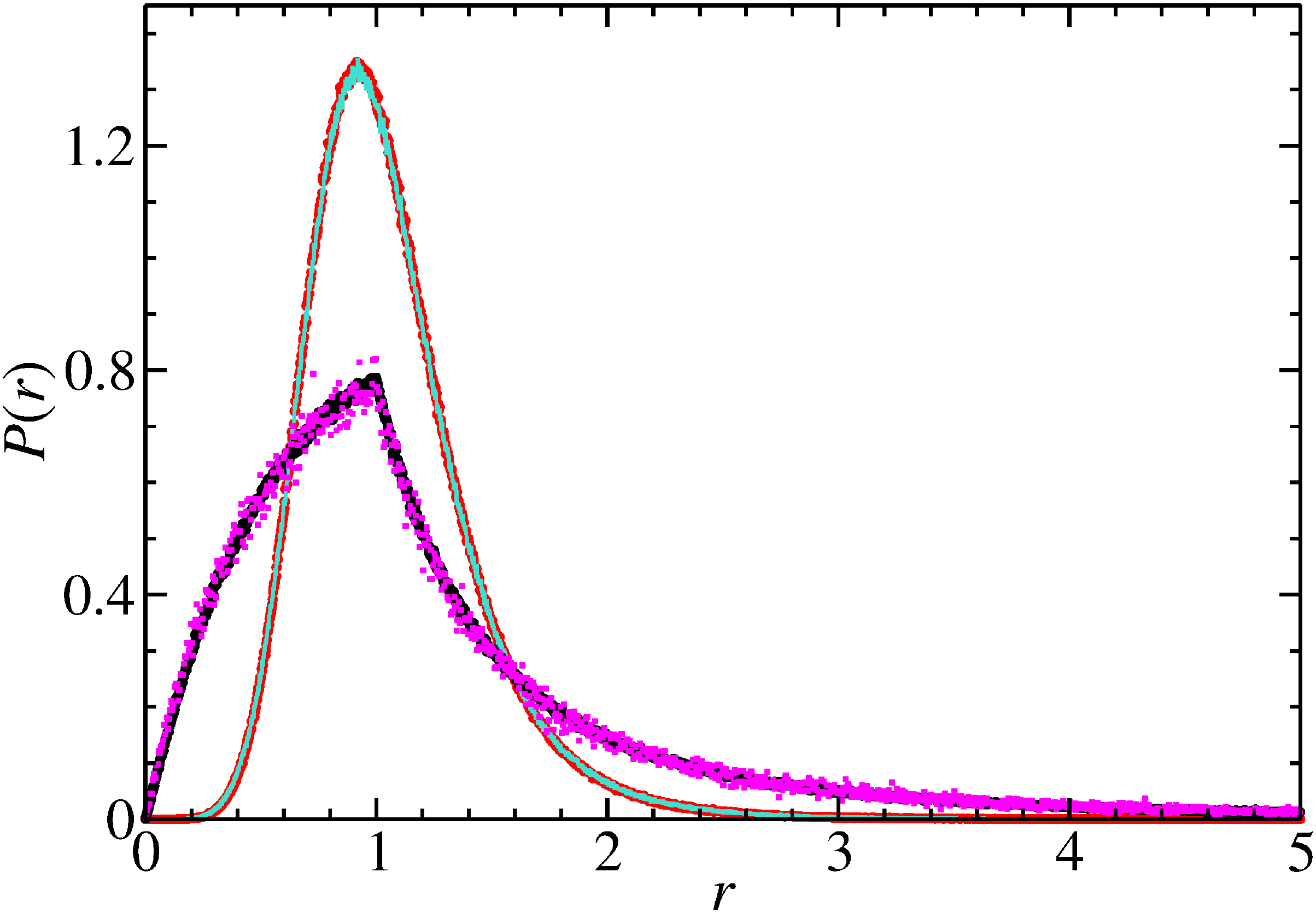}
    \caption{
        Comparison of the distributions of the ratios of the spacings betweenn next-nearest \(l=1\) (upper panel) and next-next nearest levels \(l=2\) (lower panel) for the cases A (red dots), case B (turquoise line) and C (black dots). 
        For cases A and B the curves lie on top of each other and also on top of the GUE curve, for case C they lie on top of the curve for 200000 Poissonian random numbers (magenta squares).
        }
    \label{fig:ratios}
\end{figure}

\begin{figure}[!th]
    \includegraphics[width=0.7\columnwidth]{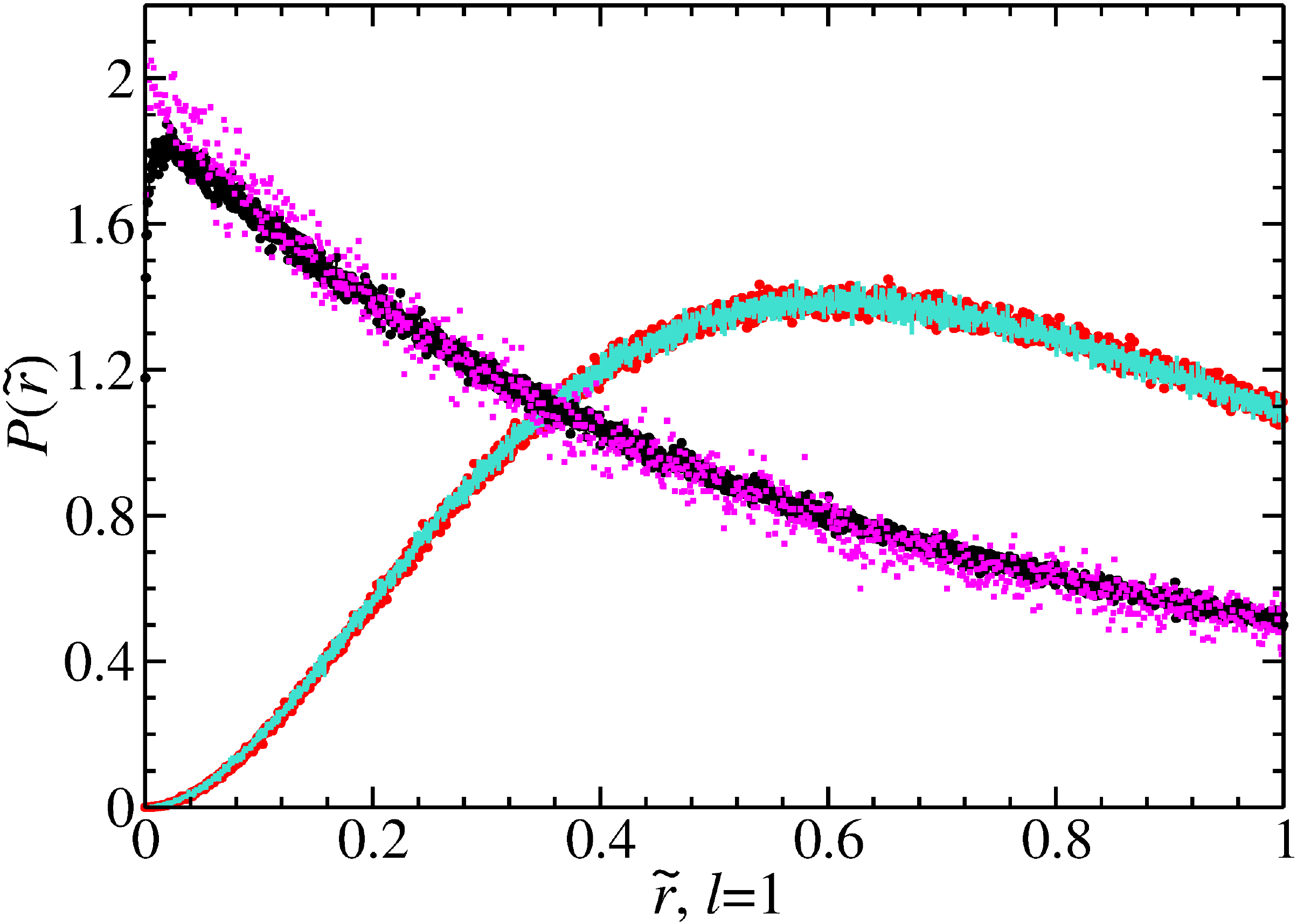}
    \includegraphics[width=0.7\columnwidth]{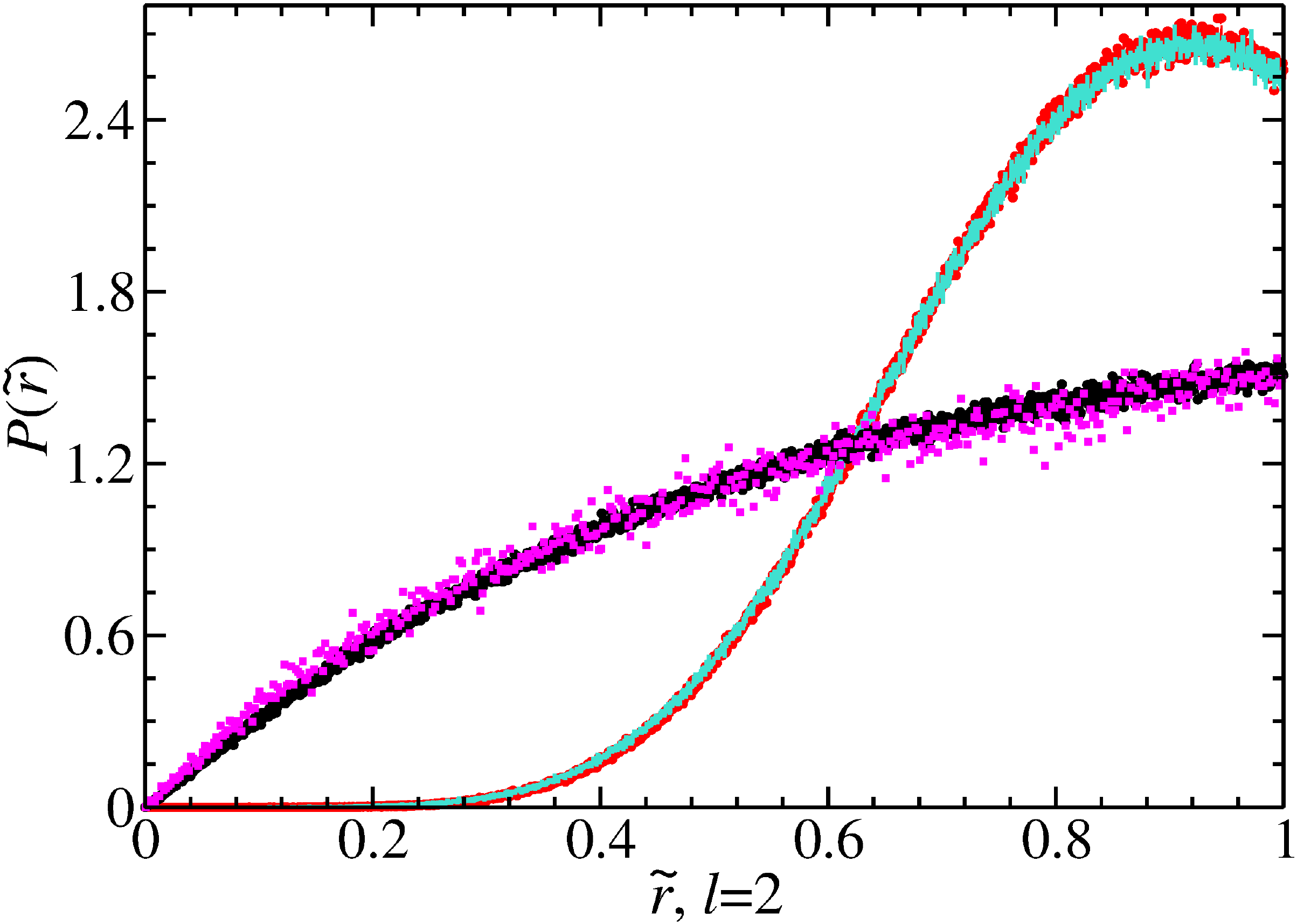}
    \caption{
        Same as~\reffig{fig:ratios} for the distribitons of \(\tilde r=\min\{r_j,\frac{1}{r_j}\}\).
    }
    \label{fig:modratios}
\end{figure}

In~\reffig{fig:spectral} we compare the spectral properties of the cases A (red histogram and circles), B (blue histogram and crosses) and C (green histogram and lines). 
They are compared with the RMT prediction for Poissonian random numbers (black dash-dotted lines) and for the GUE (black solid lines). 
For case A the curves lie on top of the GUE curves, for case C they are close to the Poisson curves. 
The short-range correlations of case B are close to those of case A, whereas we observe clear deviations for the long-range corelations. 
The same behavior is observed for the power spectra, shown in~\reffig{fig:power}.
The distributions \(r\) and \(\tilde r\), which provide another measure for short-range correlations, are exhibited in Fig.~\ref{fig:ratios} and in Fig.~\ref{fig:modratios}.
They agree well with the GUE results for cases A and B and with that of Poissonian random numbers for case C.


For case C deviations from the results for Poissonian random numbers are only observed for ratios smaller than \(0.025\), whereas the curves for cases B and C lie on top of each other and of the RMT prediction for the GUE. 
This behavior was also observed for \(l=3,\dots,10\).
It is in contrast to that observed in the spectral properties, where clear differences are observed.

\section{Details on the definition of SFF}
\label{app:sff}

Each ordered energy spectrum \(E_1 < E_2 < \dots < E_D \) is unfolded by fitting the integrated level density with a \(12\)th order polynomial.
The resulting unfolded eigenvalues are denoted by \(\epsilon_1 < \epsilon_2 < \dots < \epsilon_D\).
The Gaussian filtering function, \( \rho(\epsilon_n)\) is defined as follows
\begin{equation}
    \label{eq:gaussian_filtering_def}
    \rho (\epsilon_n) \equiv \exp{- \frac{(\epsilon_n - \bar{\epsilon})^2}{2 (\eta \Gamma)^2}} \ ,
\end{equation}
with \(\bar \epsilon\) and \( \Gamma\) being the mean value and the standard deviation, respectively, of the ensemble realization under consideration.
The parameter \( \eta\) determines the effective portion of the spectrum controlling the SFF.
It is adjusted to increase as much as possible the portion of the spectrum considered but at the same time removing the hard edges of the unfolded spectra.

The prefactor \(Z\) is chosen to make the SFF at large time equal to \(1\) and it is therefore taken equal to \( Z \equiv \langle \sum_n \abs{\rho(\epsilon_n)}^2\rangle\).


\begin{thebibliography}{62}%
\makeatletter
\providecommand \@ifxundefined [1]{%
 \@ifx{#1\undefined}
}%
\providecommand \@ifnum [1]{%
 \ifnum #1\expandafter \@firstoftwo
 \else \expandafter \@secondoftwo
 \fi
}%
\providecommand \@ifx [1]{%
 \ifx #1\expandafter \@firstoftwo
 \else \expandafter \@secondoftwo
 \fi
}%
\providecommand \natexlab [1]{#1}%
\providecommand \enquote  [1]{``#1''}%
\providecommand \bibnamefont  [1]{#1}%
\providecommand \bibfnamefont [1]{#1}%
\providecommand \citenamefont [1]{#1}%
\providecommand \href@noop [0]{\@secondoftwo}%
\providecommand \href [0]{\begingroup \@sanitize@url \@href}%
\providecommand \@href[1]{\@@startlink{#1}\@@href}%
\providecommand \@@href[1]{\endgroup#1\@@endlink}%
\providecommand \@sanitize@url [0]{\catcode `\\12\catcode `\$12\catcode
  `\&12\catcode `\#12\catcode `\^12\catcode `\_12\catcode `\%12\relax}%
\providecommand \@@startlink[1]{}%
\providecommand \@@endlink[0]{}%
\providecommand \url  [0]{\begingroup\@sanitize@url \@url }%
\providecommand \@url [1]{\endgroup\@href {#1}{\urlprefix }}%
\providecommand \urlprefix  [0]{URL }%
\providecommand \Eprint [0]{\href }%
\providecommand \doibase [0]{https://doi.org/}%
\providecommand \selectlanguage [0]{\@gobble}%
\providecommand \bibinfo  [0]{\@secondoftwo}%
\providecommand \bibfield  [0]{\@secondoftwo}%
\providecommand \translation [1]{[#1]}%
\providecommand \BibitemOpen [0]{}%
\providecommand \bibitemStop [0]{}%
\providecommand \bibitemNoStop [0]{.\EOS\space}%
\providecommand \EOS [0]{\spacefactor3000\relax}%
\providecommand \BibitemShut  [1]{\csname bibitem#1\endcsname}%
\let\auto@bib@innerbib\@empty
\bibitem [{\citenamefont {Deutsch}(1991)}]{deutsch1991quantum}%
  \BibitemOpen
  \bibfield  {author} {\bibinfo {author} {\bibfnamefont {J.~M.}\ \bibnamefont
  {Deutsch}},\ }\bibfield  {title} {\bibinfo {title} {Quantum statistical
  mechanics in a closed system},\ }\href
  {https://doi.org/10.1103/PhysRevA.43.2046} {\bibfield  {journal} {\bibinfo
  {journal} {Phys. Rev. A}\ }\textbf {\bibinfo {volume} {43}},\ \bibinfo
  {pages} {2046} (\bibinfo {year} {1991})}\BibitemShut {NoStop}%
\bibitem [{\citenamefont {Srednicki}(1994)}]{srednicki1994chaos}%
  \BibitemOpen
  \bibfield  {author} {\bibinfo {author} {\bibfnamefont {M.}~\bibnamefont
  {Srednicki}},\ }\bibfield  {title} {\bibinfo {title} {Chaos and quantum
  thermalization},\ }\href {https://doi.org/10.1103/PhysRevE.50.888} {\bibfield
   {journal} {\bibinfo  {journal} {Phys. Rev. E}\ }\textbf {\bibinfo {volume}
  {50}},\ \bibinfo {pages} {888} (\bibinfo {year} {1994})}\BibitemShut
  {NoStop}%
\bibitem [{\citenamefont {D'Alessio}\ \emph {et~al.}(2016)\citenamefont
  {D'Alessio}, \citenamefont {Kafri}, \citenamefont {Polkovnikov},\ and\
  \citenamefont {Rigol}}]{dalessio2018from}%
  \BibitemOpen
  \bibfield  {author} {\bibinfo {author} {\bibfnamefont {L.}~\bibnamefont
  {D'Alessio}}, \bibinfo {author} {\bibfnamefont {Y.}~\bibnamefont {Kafri}},
  \bibinfo {author} {\bibfnamefont {A.}~\bibnamefont {Polkovnikov}},\ and\
  \bibinfo {author} {\bibfnamefont {M.}~\bibnamefont {Rigol}},\ }\bibfield
  {title} {\bibinfo {title} {From quantum chaos and eigenstate thermalization
  to statistical mechanics and thermodynamics},\ }\href
  {https://doi.org/10.1080/00018732.2016.1198134} {\bibfield  {journal}
  {\bibinfo  {journal} {Advances in Physics}\ }\textbf {\bibinfo {volume}
  {65}},\ \bibinfo {pages} {239} (\bibinfo {year} {2016})}\BibitemShut
  {NoStop}%
\bibitem [{\citenamefont {Anderson}(1958)}]{anderson1958absence}%
  \BibitemOpen
  \bibfield  {author} {\bibinfo {author} {\bibfnamefont {P.}~\bibnamefont
  {Anderson}},\ }\bibfield  {title} {\bibinfo {title} {Absence of diffusion in
  certain random lattices},\ }\href {https://doi.org/10.1103/PhysRev.109.1492}
  {\bibfield  {journal} {\bibinfo  {journal} {Phys. Rev.}\ }\textbf {\bibinfo
  {volume} {109}},\ \bibinfo {pages} {1492} (\bibinfo {year}
  {1958})}\BibitemShut {NoStop}%
\bibitem [{\citenamefont {Kramer}\ and\ \citenamefont
  {MacKinnon}(1993)}]{kramer1993review}%
  \BibitemOpen
  \bibfield  {author} {\bibinfo {author} {\bibfnamefont {B.}~\bibnamefont
  {Kramer}}\ and\ \bibinfo {author} {\bibfnamefont {A.}~\bibnamefont
  {MacKinnon}},\ }\bibfield  {title} {\bibinfo {title} {Localization: theory
  and experiment},\ }\href {https://doi.org/10.1088/0034-4885/56/12/001}
  {\bibfield  {journal} {\bibinfo  {journal} {Reports on Progress in Physics}\
  }\textbf {\bibinfo {volume} {56}},\ \bibinfo {pages} {1469} (\bibinfo {year}
  {1993})}\BibitemShut {NoStop}%
\bibitem [{\citenamefont {Evers}\ and\ \citenamefont
  {Mirlin}(2008)}]{evers2008anderson}%
  \BibitemOpen
  \bibfield  {author} {\bibinfo {author} {\bibfnamefont {F.}~\bibnamefont
  {Evers}}\ and\ \bibinfo {author} {\bibfnamefont {A.~D.}\ \bibnamefont
  {Mirlin}},\ }\bibfield  {title} {\bibinfo {title} {Anderson transitions},\
  }\href {https://doi.org/10.1103/RevModPhys.80.1355} {\bibfield  {journal}
  {\bibinfo  {journal} {Rev. Mod. Phys.}\ }\textbf {\bibinfo {volume} {80}},\
  \bibinfo {pages} {1355} (\bibinfo {year} {2008})}\BibitemShut {NoStop}%
\bibitem [{\citenamefont {Hundertmark}(2008)}]{hundertmark2008a}%
  \BibitemOpen
  \bibfield  {author} {\bibinfo {author} {\bibfnamefont {D.}~\bibnamefont
  {Hundertmark}},\ }\bibfield  {title} {\bibinfo {title} {A short introduction
  to anderson localization},\ }\href@noop {} {\bibfield  {journal} {\bibinfo
  {journal} {Analysis and Stochastics of Growth Processes and Interface
  Models}\ ,\ \bibinfo {pages} {194}} (\bibinfo {year} {2008})}\BibitemShut
  {NoStop}%
\bibitem [{\citenamefont {Thouless}(1972)}]{thouless1972a}%
  \BibitemOpen
  \bibfield  {author} {\bibinfo {author} {\bibfnamefont {D.~J.}\ \bibnamefont
  {Thouless}},\ }\bibfield  {title} {\bibinfo {title} {A relation between the
  density of states and range of localization for one dimensional random
  systems},\ }\href {https://doi.org/10.1088/0022-3719/5/1/010} {\bibfield
  {journal} {\bibinfo  {journal} {Journal of Physics C: Solid State Physics}\
  }\textbf {\bibinfo {volume} {5}},\ \bibinfo {pages} {77} (\bibinfo {year}
  {1972})}\BibitemShut {NoStop}%
\bibitem [{\citenamefont {Anderson}\ \emph {et~al.}(1980)\citenamefont
  {Anderson}, \citenamefont {Thouless}, \citenamefont {Abrahams},\ and\
  \citenamefont {Fisher}}]{anderson1980new}%
  \BibitemOpen
  \bibfield  {author} {\bibinfo {author} {\bibfnamefont {P.~W.}\ \bibnamefont
  {Anderson}}, \bibinfo {author} {\bibfnamefont {D.~J.}\ \bibnamefont
  {Thouless}}, \bibinfo {author} {\bibfnamefont {E.}~\bibnamefont {Abrahams}},\
  and\ \bibinfo {author} {\bibfnamefont {D.~S.}\ \bibnamefont {Fisher}},\
  }\bibfield  {title} {\bibinfo {title} {New method for a scaling theory of
  localization},\ }\href {https://doi.org/10.1103/PhysRevB.22.3519} {\bibfield
  {journal} {\bibinfo  {journal} {Phys. Rev. B}\ }\textbf {\bibinfo {volume}
  {22}},\ \bibinfo {pages} {3519} (\bibinfo {year} {1980})}\BibitemShut
  {NoStop}%
\bibitem [{\citenamefont {Bardarson}\ \emph {et~al.}(2012)\citenamefont
  {Bardarson}, \citenamefont {Pollmann},\ and\ \citenamefont
  {Moore}}]{bardarson2012unbounded}%
  \BibitemOpen
  \bibfield  {author} {\bibinfo {author} {\bibfnamefont {J.~H.}\ \bibnamefont
  {Bardarson}}, \bibinfo {author} {\bibfnamefont {F.}~\bibnamefont
  {Pollmann}},\ and\ \bibinfo {author} {\bibfnamefont {J.~E.}\ \bibnamefont
  {Moore}},\ }\bibfield  {title} {\bibinfo {title} {Unbounded growth of
  entanglement in models of many-body localization},\ }\href
  {https://doi.org/10.1103/PhysRevLett.109.017202} {\bibfield  {journal}
  {\bibinfo  {journal} {Phys. Rev. Lett.}\ }\textbf {\bibinfo {volume} {109}},\
  \bibinfo {pages} {017202} (\bibinfo {year} {2012})}\BibitemShut {NoStop}%
\bibitem [{\citenamefont {Altshuler}\ \emph {et~al.}(1997)\citenamefont
  {Altshuler}, \citenamefont {Gefen}, \citenamefont {Kamenev},\ and\
  \citenamefont {Levitov}}]{altshuler1997quasi}%
  \BibitemOpen
  \bibfield  {author} {\bibinfo {author} {\bibfnamefont {B.~L.}\ \bibnamefont
  {Altshuler}}, \bibinfo {author} {\bibfnamefont {Y.}~\bibnamefont {Gefen}},
  \bibinfo {author} {\bibfnamefont {A.}~\bibnamefont {Kamenev}},\ and\ \bibinfo
  {author} {\bibfnamefont {L.~S.}\ \bibnamefont {Levitov}},\ }\bibfield
  {title} {\bibinfo {title} {Quasiparticle lifetime in a finite system: A
  nonperturbative approach},\ }\href
  {https://doi.org/10.1103/PhysRevLett.78.2803} {\bibfield  {journal} {\bibinfo
   {journal} {Phys. Rev. Lett.}\ }\textbf {\bibinfo {volume} {78}},\ \bibinfo
  {pages} {2803} (\bibinfo {year} {1997})}\BibitemShut {NoStop}%
\bibitem [{\citenamefont {Gornyi}\ \emph {et~al.}(2005)\citenamefont {Gornyi},
  \citenamefont {Mirlin},\ and\ \citenamefont
  {Polyakov}}]{gornyi2005interacting}%
  \BibitemOpen
  \bibfield  {author} {\bibinfo {author} {\bibfnamefont {I.}~\bibnamefont
  {Gornyi}}, \bibinfo {author} {\bibfnamefont {A.}~\bibnamefont {Mirlin}},\
  and\ \bibinfo {author} {\bibfnamefont {D.}~\bibnamefont {Polyakov}},\
  }\bibfield  {title} {\bibinfo {title} {Interacting electrons in disordered
  wires: Anderson localization and low-$t$ transport},\ }\href
  {https://doi.org/10.1103/PhysRevLett.95.206603} {\bibfield  {journal}
  {\bibinfo  {journal} {Phys. Rev. Lett.}\ }\textbf {\bibinfo {volume} {95}},\
  \bibinfo {pages} {206603} (\bibinfo {year} {2005})}\BibitemShut {NoStop}%
\bibitem [{\citenamefont {Basko}\ \emph {et~al.}(2006)\citenamefont {Basko},
  \citenamefont {Aleiner},\ and\ \citenamefont {Altshuler}}]{basko2006metal}%
  \BibitemOpen
  \bibfield  {author} {\bibinfo {author} {\bibfnamefont {D.}~\bibnamefont
  {Basko}}, \bibinfo {author} {\bibfnamefont {I.}~\bibnamefont {Aleiner}},\
  and\ \bibinfo {author} {\bibfnamefont {B.}~\bibnamefont {Altshuler}},\
  }\bibfield  {title} {\bibinfo {title} {Metal--insulator transition in a
  weakly interacting many-electron system with localized single-particle
  states},\ }\href {https://doi.org/10.1016/j.aop.2005.11.014} {\bibfield
  {journal} {\bibinfo  {journal} {Ann. Phys.}\ }\textbf {\bibinfo {volume}
  {321}},\ \bibinfo {pages} {1126 } (\bibinfo {year} {2006})}\BibitemShut
  {NoStop}%
\bibitem [{\citenamefont {Fleishman}\ and\ \citenamefont
  {Anderson}(1980)}]{fleishman1980interaction}%
  \BibitemOpen
  \bibfield  {author} {\bibinfo {author} {\bibfnamefont {L.}~\bibnamefont
  {Fleishman}}\ and\ \bibinfo {author} {\bibfnamefont {P.}~\bibnamefont
  {Anderson}},\ }\bibfield  {title} {\bibinfo {title} {Interactions and the
  anderson transition},\ }\href {https://doi.org/10.1103/PhysRevB.21.2366}
  {\bibfield  {journal} {\bibinfo  {journal} {Phys. Rev. B}\ }\textbf {\bibinfo
  {volume} {21}},\ \bibinfo {pages} {2366} (\bibinfo {year}
  {1980})}\BibitemShut {NoStop}%
\bibitem [{\citenamefont {Serbyn}\ \emph {et~al.}(2013)\citenamefont {Serbyn},
  \citenamefont {Papi\ifmmode~\acute{c}\else \'{c}\fi{}},\ and\ \citenamefont
  {Abanin}}]{serbyn2013local}%
  \BibitemOpen
  \bibfield  {author} {\bibinfo {author} {\bibfnamefont {M.}~\bibnamefont
  {Serbyn}}, \bibinfo {author} {\bibfnamefont {Z.}~\bibnamefont
  {Papi\ifmmode~\acute{c}\else \'{c}\fi{}}},\ and\ \bibinfo {author}
  {\bibfnamefont {D.~A.}\ \bibnamefont {Abanin}},\ }\bibfield  {title}
  {\bibinfo {title} {Local conservation laws and the structure of the many-body
  localized states},\ }\href {https://doi.org/10.1103/PhysRevLett.111.127201}
  {\bibfield  {journal} {\bibinfo  {journal} {Phys. Rev. Lett.}\ }\textbf
  {\bibinfo {volume} {111}},\ \bibinfo {pages} {127201} (\bibinfo {year}
  {2013})}\BibitemShut {NoStop}%
\bibitem [{\citenamefont {Bauer}\ and\ \citenamefont
  {Nayak}(2013)}]{bauer2013area}%
  \BibitemOpen
  \bibfield  {author} {\bibinfo {author} {\bibfnamefont {B.}~\bibnamefont
  {Bauer}}\ and\ \bibinfo {author} {\bibfnamefont {C.}~\bibnamefont {Nayak}},\
  }\bibfield  {title} {\bibinfo {title} {Area laws in a many-body localized
  state and its implications for topological order},\ }\href
  {https://doi.org/10.1088/1742-5468/2013/09/p09005} {\bibfield  {journal}
  {\bibinfo  {journal} {Journal of Statistical Mechanics: Theory and
  Experiment}\ }\textbf {\bibinfo {volume} {2013}},\ \bibinfo {pages} {P09005}
  (\bibinfo {year} {2013})}\BibitemShut {NoStop}%
\bibitem [{\citenamefont {\ifmmode \check{Z}\else
  \v{Z}\fi{}nidari\ifmmode~\check{c}\else \v{c}\fi{}}\ \emph
  {et~al.}(2008)\citenamefont {\ifmmode \check{Z}\else
  \v{Z}\fi{}nidari\ifmmode~\check{c}\else \v{c}\fi{}}, \citenamefont {Prosen},\
  and\ \citenamefont {Prelov\ifmmode~\check{s}\else
  \v{s}\fi{}ek}}]{znidaric2008many-body}%
  \BibitemOpen
  \bibfield  {author} {\bibinfo {author} {\bibfnamefont {M.}~\bibnamefont
  {\ifmmode \check{Z}\else \v{Z}\fi{}nidari\ifmmode~\check{c}\else
  \v{c}\fi{}}}, \bibinfo {author} {\bibfnamefont {T.}~\bibnamefont {Prosen}},\
  and\ \bibinfo {author} {\bibfnamefont {P.}~\bibnamefont
  {Prelov\ifmmode~\check{s}\else \v{s}\fi{}ek}},\ }\bibfield  {title} {\bibinfo
  {title} {Many-body localization in the heisenberg $xxz$ magnet in a random
  field},\ }\href {https://doi.org/10.1103/PhysRevB.77.064426} {\bibfield
  {journal} {\bibinfo  {journal} {Phys. Rev. B}\ }\textbf {\bibinfo {volume}
  {77}},\ \bibinfo {pages} {064426} (\bibinfo {year} {2008})}\BibitemShut
  {NoStop}%
\bibitem [{\citenamefont {Oganesyan}\ and\ \citenamefont
  {Huse}(2007)}]{oganesyan2007localization}%
  \BibitemOpen
  \bibfield  {author} {\bibinfo {author} {\bibfnamefont {V.}~\bibnamefont
  {Oganesyan}}\ and\ \bibinfo {author} {\bibfnamefont {D.~A.}\ \bibnamefont
  {Huse}},\ }\bibfield  {title} {\bibinfo {title} {Localization of interacting
  fermions at high temperature},\ }\href
  {https://doi.org/10.1103/PhysRevB.75.155111} {\bibfield  {journal} {\bibinfo
  {journal} {Phys. Rev. B}\ }\textbf {\bibinfo {volume} {75}},\ \bibinfo
  {pages} {155111} (\bibinfo {year} {2007})}\BibitemShut {NoStop}%
\bibitem [{\citenamefont {Pal}\ and\ \citenamefont
  {Huse}(2010)}]{pal2010many-body}%
  \BibitemOpen
  \bibfield  {author} {\bibinfo {author} {\bibfnamefont {A.}~\bibnamefont
  {Pal}}\ and\ \bibinfo {author} {\bibfnamefont {D.~A.}\ \bibnamefont {Huse}},\
  }\bibfield  {title} {\bibinfo {title} {Many-body localization phase
  transition},\ }\href {https://doi.org/10.1103/PhysRevB.82.174411} {\bibfield
  {journal} {\bibinfo  {journal} {Phys. Rev. B}\ }\textbf {\bibinfo {volume}
  {82}},\ \bibinfo {pages} {174411} (\bibinfo {year} {2010})}\BibitemShut
  {NoStop}%
\bibitem [{\citenamefont {Ros}\ \emph {et~al.}(2015)\citenamefont {Ros},
  \citenamefont {Müller},\ and\ \citenamefont
  {Scardicchio}}]{ros2015integrals}%
  \BibitemOpen
  \bibfield  {author} {\bibinfo {author} {\bibfnamefont {V.}~\bibnamefont
  {Ros}}, \bibinfo {author} {\bibfnamefont {M.}~\bibnamefont {Müller}},\ and\
  \bibinfo {author} {\bibfnamefont {A.}~\bibnamefont {Scardicchio}},\
  }\bibfield  {title} {\bibinfo {title} {Integrals of motion in the many-body
  localized phase},\ }\href
  {https://doi.org/https://doi.org/10.1016/j.nuclphysb.2014.12.014} {\bibfield
  {journal} {\bibinfo  {journal} {Nucl. Phys. B}\ }\textbf {\bibinfo {volume}
  {891}},\ \bibinfo {pages} {420 } (\bibinfo {year} {2015})}\BibitemShut
  {NoStop}%
\bibitem [{\citenamefont {Huse}\ \emph {et~al.}(2014)\citenamefont {Huse},
  \citenamefont {Nandkishore},\ and\ \citenamefont
  {Oganesyan}}]{huse2014phenomenology}%
  \BibitemOpen
  \bibfield  {author} {\bibinfo {author} {\bibfnamefont {D.~A.}\ \bibnamefont
  {Huse}}, \bibinfo {author} {\bibfnamefont {R.}~\bibnamefont {Nandkishore}},\
  and\ \bibinfo {author} {\bibfnamefont {V.}~\bibnamefont {Oganesyan}},\
  }\bibfield  {title} {\bibinfo {title} {Phenomenology of fully
  many-body-localized systems},\ }\href
  {https://doi.org/10.1103/PhysRevB.90.174202} {\bibfield  {journal} {\bibinfo
  {journal} {Phys. Rev. B}\ }\textbf {\bibinfo {volume} {90}},\ \bibinfo
  {pages} {174202} (\bibinfo {year} {2014})}\BibitemShut {NoStop}%
\bibitem [{\citenamefont {Luitz}\ \emph {et~al.}(2015)\citenamefont {Luitz},
  \citenamefont {Laflorencie},\ and\ \citenamefont {Alet}}]{luitz2015many}%
  \BibitemOpen
  \bibfield  {author} {\bibinfo {author} {\bibfnamefont {D.~J.}\ \bibnamefont
  {Luitz}}, \bibinfo {author} {\bibfnamefont {N.}~\bibnamefont {Laflorencie}},\
  and\ \bibinfo {author} {\bibfnamefont {F.}~\bibnamefont {Alet}},\ }\bibfield
  {title} {\bibinfo {title} {Many-body localization edge in the random-field
  heisenberg chain},\ }\href {https://doi.org/10.1103/PhysRevB.91.081103}
  {\bibfield  {journal} {\bibinfo  {journal} {Phys. Rev. B}\ }\textbf {\bibinfo
  {volume} {91}},\ \bibinfo {pages} {081103} (\bibinfo {year}
  {2015})}\BibitemShut {NoStop}%
\bibitem [{\citenamefont {\ifmmode~\check{S}\else \v{S}\fi{}untajs}\ \emph
  {et~al.}(2020)\citenamefont {\ifmmode~\check{S}\else \v{S}\fi{}untajs},
  \citenamefont {Bon\ifmmode~\check{c}\else \v{c}\fi{}a}, \citenamefont
  {Prosen},\ and\ \citenamefont {Vidmar}}]{suntas2020quantum}%
  \BibitemOpen
  \bibfield  {author} {\bibinfo {author} {\bibfnamefont {J.}~\bibnamefont
  {\ifmmode~\check{S}\else \v{S}\fi{}untajs}}, \bibinfo {author} {\bibfnamefont
  {J.}~\bibnamefont {Bon\ifmmode~\check{c}\else \v{c}\fi{}a}}, \bibinfo
  {author} {\bibfnamefont {T.}~\bibnamefont {Prosen}},\ and\ \bibinfo {author}
  {\bibfnamefont {L.}~\bibnamefont {Vidmar}},\ }\bibfield  {title} {\bibinfo
  {title} {Quantum chaos challenges many-body localization},\ }\href
  {https://doi.org/10.1103/PhysRevE.102.062144} {\bibfield  {journal} {\bibinfo
   {journal} {Phys. Rev. E}\ }\textbf {\bibinfo {volume} {102}},\ \bibinfo
  {pages} {062144} (\bibinfo {year} {2020})}\BibitemShut {NoStop}%
\bibitem [{\citenamefont {\v{S}untajs}\ \emph {et~al.}(2020)\citenamefont
  {\v{S}untajs}, \citenamefont {Bon\v{c}a}, \citenamefont {Prosen},\ and\
  \citenamefont {Vidmar}}]{suntas2020ergodicity}%
  \BibitemOpen
  \bibfield  {author} {\bibinfo {author} {\bibfnamefont {J.}~\bibnamefont
  {\v{S}untajs}}, \bibinfo {author} {\bibfnamefont {J.}~\bibnamefont
  {Bon\v{c}a}}, \bibinfo {author} {\bibfnamefont {T.}~\bibnamefont {Prosen}},\
  and\ \bibinfo {author} {\bibfnamefont {L.}~\bibnamefont {Vidmar}},\
  }\bibfield  {title} {\bibinfo {title} {Ergodicity breaking transition in
  finite disordered spin chains},\ }\href
  {https://doi.org/10.1103/PhysRevB.102.064207} {\bibfield  {journal} {\bibinfo
   {journal} {Phys. Rev. B}\ }\textbf {\bibinfo {volume} {102}},\ \bibinfo
  {pages} {064207} (\bibinfo {year} {2020})}\BibitemShut {NoStop}%
\bibitem [{\citenamefont {Abanin}\ \emph {et~al.}(2021)\citenamefont {Abanin},
  \citenamefont {Bardarson}, \citenamefont {{De Tomasi}}, \citenamefont
  {Gopalakrishnan}, \citenamefont {Khemani}, \citenamefont {Parameswaran},
  \citenamefont {Pollmann}, \citenamefont {Potter}, \citenamefont {Serbyn},\
  and\ \citenamefont {Vasseur}}]{abanin2021distinguishing}%
  \BibitemOpen
  \bibfield  {author} {\bibinfo {author} {\bibfnamefont {D.}~\bibnamefont
  {Abanin}}, \bibinfo {author} {\bibfnamefont {J.}~\bibnamefont {Bardarson}},
  \bibinfo {author} {\bibfnamefont {G.}~\bibnamefont {{De Tomasi}}}, \bibinfo
  {author} {\bibfnamefont {S.}~\bibnamefont {Gopalakrishnan}}, \bibinfo
  {author} {\bibfnamefont {V.}~\bibnamefont {Khemani}}, \bibinfo {author}
  {\bibfnamefont {S.}~\bibnamefont {Parameswaran}}, \bibinfo {author}
  {\bibfnamefont {F.}~\bibnamefont {Pollmann}}, \bibinfo {author}
  {\bibfnamefont {A.}~\bibnamefont {Potter}}, \bibinfo {author} {\bibfnamefont
  {M.}~\bibnamefont {Serbyn}},\ and\ \bibinfo {author} {\bibfnamefont
  {R.}~\bibnamefont {Vasseur}},\ }\bibfield  {title} {\bibinfo {title}
  {Distinguishing localization from chaos: Challenges in finite-size systems},\
  }\href {https://doi.org/https://doi.org/10.1016/j.aop.2021.168415} {\bibfield
   {journal} {\bibinfo  {journal} {Annals of Physics}\ }\textbf {\bibinfo
  {volume} {427}},\ \bibinfo {pages} {168415} (\bibinfo {year}
  {2021})}\BibitemShut {NoStop}%
\bibitem [{\citenamefont {Sels}\ and\ \citenamefont
  {Polkovnikov}(2021)}]{sels2021dynamical}%
  \BibitemOpen
  \bibfield  {author} {\bibinfo {author} {\bibfnamefont {D.}~\bibnamefont
  {Sels}}\ and\ \bibinfo {author} {\bibfnamefont {A.}~\bibnamefont
  {Polkovnikov}},\ }\bibfield  {title} {\bibinfo {title} {Dynamical obstruction
  to localization in a disordered spin chain},\ }\href
  {https://doi.org/10.1103/PhysRevE.104.054105} {\bibfield  {journal} {\bibinfo
   {journal} {Phys. Rev. E}\ }\textbf {\bibinfo {volume} {104}},\ \bibinfo
  {pages} {054105} (\bibinfo {year} {2021})}\BibitemShut {NoStop}%
\bibitem [{\citenamefont {Morningstar}\ \emph {et~al.}(2022)\citenamefont
  {Morningstar}, \citenamefont {Colmenarez}, \citenamefont {Khemani},
  \citenamefont {Luitz},\ and\ \citenamefont
  {Huse}}]{morningstar2022avalanches}%
  \BibitemOpen
  \bibfield  {author} {\bibinfo {author} {\bibfnamefont {A.}~\bibnamefont
  {Morningstar}}, \bibinfo {author} {\bibfnamefont {L.}~\bibnamefont
  {Colmenarez}}, \bibinfo {author} {\bibfnamefont {V.}~\bibnamefont {Khemani}},
  \bibinfo {author} {\bibfnamefont {D.~J.}\ \bibnamefont {Luitz}},\ and\
  \bibinfo {author} {\bibfnamefont {D.~A.}\ \bibnamefont {Huse}},\ }\bibfield
  {title} {\bibinfo {title} {Avalanches and many-body resonances in many-body
  localized systems},\ }\href {https://doi.org/10.1103/PhysRevB.105.174205}
  {\bibfield  {journal} {\bibinfo  {journal} {Phys. Rev. B}\ }\textbf {\bibinfo
  {volume} {105}},\ \bibinfo {pages} {174205} (\bibinfo {year}
  {2022})}\BibitemShut {NoStop}%
\bibitem [{\citenamefont {Garc\'{\i}a-Garc\'{\i}a}\ \emph
  {et~al.}(2018{\natexlab{a}})\citenamefont {Garc\'{\i}a-Garc\'{\i}a},
  \citenamefont {Loureiro}, \citenamefont {Romero-Berm\'udez},\ and\
  \citenamefont {Tezuka}}]{garcia-garcia2018chaotic}%
  \BibitemOpen
  \bibfield  {author} {\bibinfo {author} {\bibfnamefont {A.~M.}\ \bibnamefont
  {Garc\'{\i}a-Garc\'{\i}a}}, \bibinfo {author} {\bibfnamefont
  {B.}~\bibnamefont {Loureiro}}, \bibinfo {author} {\bibfnamefont
  {A.}~\bibnamefont {Romero-Berm\'udez}},\ and\ \bibinfo {author}
  {\bibfnamefont {M.}~\bibnamefont {Tezuka}},\ }\bibfield  {title} {\bibinfo
  {title} {Chaotic-integrable transition in the sachdev-ye-kitaev model},\
  }\href {https://doi.org/10.1103/PhysRevLett.120.241603} {\bibfield  {journal}
  {\bibinfo  {journal} {Phys. Rev. Lett.}\ }\textbf {\bibinfo {volume} {120}},\
  \bibinfo {pages} {241603} (\bibinfo {year} {2018}{\natexlab{a}})}\BibitemShut
  {NoStop}%
\bibitem [{\citenamefont {{Nosaka}}\ \emph {et~al.}(2018)\citenamefont
  {{Nosaka}}, \citenamefont {{Rosa}},\ and\ \citenamefont
  {{Yoon}}}]{nosaka2018the}%
  \BibitemOpen
  \bibfield  {author} {\bibinfo {author} {\bibfnamefont {T.}~\bibnamefont
  {{Nosaka}}}, \bibinfo {author} {\bibfnamefont {D.}~\bibnamefont {{Rosa}}},\
  and\ \bibinfo {author} {\bibfnamefont {J.}~\bibnamefont {{Yoon}}},\
  }\bibfield  {title} {\bibinfo {title} {{The Thouless time for mass-deformed
  SYK}},\ }\href {https://doi.org/10.1007/JHEP09(2018)041} {\bibfield
  {journal} {\bibinfo  {journal} {Journal of High Energy Physics}\ }\textbf
  {\bibinfo {volume} {2018}},\ \bibinfo {eid} {41} (\bibinfo {year}
  {2018})}\BibitemShut {NoStop}%
\bibitem [{\citenamefont {Monteiro}\ \emph
  {et~al.}(2021{\natexlab{a}})\citenamefont {Monteiro}, \citenamefont
  {Micklitz}, \citenamefont {Tezuka},\ and\ \citenamefont
  {Altland}}]{monteiro2021minimal}%
  \BibitemOpen
  \bibfield  {author} {\bibinfo {author} {\bibfnamefont {F.}~\bibnamefont
  {Monteiro}}, \bibinfo {author} {\bibfnamefont {T.}~\bibnamefont {Micklitz}},
  \bibinfo {author} {\bibfnamefont {M.}~\bibnamefont {Tezuka}},\ and\ \bibinfo
  {author} {\bibfnamefont {A.}~\bibnamefont {Altland}},\ }\bibfield  {title}
  {\bibinfo {title} {Minimal model of many-body localization},\ }\href
  {https://doi.org/10.1103/PhysRevResearch.3.013023} {\bibfield  {journal}
  {\bibinfo  {journal} {Phys. Rev. Research}\ }\textbf {\bibinfo {volume}
  {3}},\ \bibinfo {pages} {013023} (\bibinfo {year}
  {2021}{\natexlab{a}})}\BibitemShut {NoStop}%
\bibitem [{\citenamefont {Monteiro}\ \emph
  {et~al.}(2021{\natexlab{b}})\citenamefont {Monteiro}, \citenamefont {Tezuka},
  \citenamefont {Altland}, \citenamefont {Huse},\ and\ \citenamefont
  {Micklitz}}]{monteiro2021quantum}%
  \BibitemOpen
  \bibfield  {author} {\bibinfo {author} {\bibfnamefont {F.}~\bibnamefont
  {Monteiro}}, \bibinfo {author} {\bibfnamefont {M.}~\bibnamefont {Tezuka}},
  \bibinfo {author} {\bibfnamefont {A.}~\bibnamefont {Altland}}, \bibinfo
  {author} {\bibfnamefont {D.~A.}\ \bibnamefont {Huse}},\ and\ \bibinfo
  {author} {\bibfnamefont {T.}~\bibnamefont {Micklitz}},\ }\bibfield  {title}
  {\bibinfo {title} {Quantum ergodicity in the many-body localization
  problem},\ }\href {https://doi.org/10.1103/PhysRevLett.127.030601} {\bibfield
   {journal} {\bibinfo  {journal} {Phys. Rev. Lett.}\ }\textbf {\bibinfo
  {volume} {127}},\ \bibinfo {pages} {030601} (\bibinfo {year}
  {2021}{\natexlab{b}})}\BibitemShut {NoStop}%
\bibitem [{\citenamefont {Pandey}\ \emph {et~al.}(2020)\citenamefont {Pandey},
  \citenamefont {Claeys}, \citenamefont {Campbell}, \citenamefont
  {Polkovnikov},\ and\ \citenamefont {Sels}}]{pandey2020adiabatic}%
  \BibitemOpen
  \bibfield  {author} {\bibinfo {author} {\bibfnamefont {M.}~\bibnamefont
  {Pandey}}, \bibinfo {author} {\bibfnamefont {P.~W.}\ \bibnamefont {Claeys}},
  \bibinfo {author} {\bibfnamefont {D.~K.}\ \bibnamefont {Campbell}}, \bibinfo
  {author} {\bibfnamefont {A.}~\bibnamefont {Polkovnikov}},\ and\ \bibinfo
  {author} {\bibfnamefont {D.}~\bibnamefont {Sels}},\ }\bibfield  {title}
  {\bibinfo {title} {Adiabatic eigenstate deformations as a sensitive probe for
  quantum chaos},\ }\href {https://doi.org/10.1103/PhysRevX.10.041017}
  {\bibfield  {journal} {\bibinfo  {journal} {Phys. Rev. X}\ }\textbf {\bibinfo
  {volume} {10}},\ \bibinfo {pages} {041017} (\bibinfo {year}
  {2020})}\BibitemShut {NoStop}%
\bibitem [{\citenamefont {Kim}\ and\ \citenamefont
  {Cao}(2021)}]{kim2021comment}%
  \BibitemOpen
  \bibfield  {author} {\bibinfo {author} {\bibfnamefont {J.}~\bibnamefont
  {Kim}}\ and\ \bibinfo {author} {\bibfnamefont {X.}~\bibnamefont {Cao}},\
  }\bibfield  {title} {\bibinfo {title} {Comment on ``chaotic-integrable
  transition in the sachdev-ye-kitaev model''},\ }\href
  {https://doi.org/10.1103/PhysRevLett.126.109101} {\bibfield  {journal}
  {\bibinfo  {journal} {Phys. Rev. Lett.}\ }\textbf {\bibinfo {volume} {126}},\
  \bibinfo {pages} {109101} (\bibinfo {year} {2021})}\BibitemShut {NoStop}%
\bibitem [{\citenamefont {Garc\'{\i}a-Garc\'{\i}a}\ \emph
  {et~al.}(2021)\citenamefont {Garc\'{\i}a-Garc\'{\i}a}, \citenamefont
  {Loureiro}, \citenamefont {Romero-Berm\'udez},\ and\ \citenamefont
  {Tezuka}}]{garcia-garcia2021reply}%
  \BibitemOpen
  \bibfield  {author} {\bibinfo {author} {\bibfnamefont {A.~M.}\ \bibnamefont
  {Garc\'{\i}a-Garc\'{\i}a}}, \bibinfo {author} {\bibfnamefont
  {B.}~\bibnamefont {Loureiro}}, \bibinfo {author} {\bibfnamefont
  {A.}~\bibnamefont {Romero-Berm\'udez}},\ and\ \bibinfo {author}
  {\bibfnamefont {M.}~\bibnamefont {Tezuka}},\ }\bibfield  {title} {\bibinfo
  {title} {Garc\'{\i}a-garc\'{\i}a et al. reply:},\ }\href
  {https://doi.org/10.1103/PhysRevLett.126.109102} {\bibfield  {journal}
  {\bibinfo  {journal} {Phys. Rev. Lett.}\ }\textbf {\bibinfo {volume} {126}},\
  \bibinfo {pages} {109102} (\bibinfo {year} {2021})}\BibitemShut {NoStop}%
\bibitem [{\citenamefont {Sachdev}\ and\ \citenamefont
  {Ye}(1993)}]{sachdev1993spin}%
  \BibitemOpen
  \bibfield  {author} {\bibinfo {author} {\bibfnamefont {S.}~\bibnamefont
  {Sachdev}}\ and\ \bibinfo {author} {\bibfnamefont {J.}~\bibnamefont {Ye}},\
  }\bibfield  {title} {\bibinfo {title} {Gapless spin-fluid ground state in a
  random quantum heisenberg magnet},\ }\href
  {https://doi.org/10.1103/PhysRevLett.70.3339} {\bibfield  {journal} {\bibinfo
   {journal} {Phys. Rev. Lett.}\ }\textbf {\bibinfo {volume} {70}},\ \bibinfo
  {pages} {3339} (\bibinfo {year} {1993})}\BibitemShut {NoStop}%
\bibitem [{\citenamefont {Maldacena}\ and\ \citenamefont
  {Stanford}(2016)}]{maldacena2016remarks}%
  \BibitemOpen
  \bibfield  {author} {\bibinfo {author} {\bibfnamefont {J.}~\bibnamefont
  {Maldacena}}\ and\ \bibinfo {author} {\bibfnamefont {D.}~\bibnamefont
  {Stanford}},\ }\bibfield  {title} {\bibinfo {title} {Remarks on the
  sachdev-ye-kitaev model},\ }\href
  {https://doi.org/10.1103/PhysRevD.94.106002} {\bibfield  {journal} {\bibinfo
  {journal} {Phys. Rev. D}\ }\textbf {\bibinfo {volume} {94}},\ \bibinfo
  {pages} {106002} (\bibinfo {year} {2016})}\BibitemShut {NoStop}%
\bibitem [{\citenamefont {Polchinski}\ and\ \citenamefont
  {Rosenhaus}(2016)}]{polchinski2016the}%
  \BibitemOpen
  \bibfield  {author} {\bibinfo {author} {\bibfnamefont {J.}~\bibnamefont
  {Polchinski}}\ and\ \bibinfo {author} {\bibfnamefont {V.}~\bibnamefont
  {Rosenhaus}},\ }\bibfield  {title} {\bibinfo {title} {The spectrum in the
  sachdev-ye-kitaev model},\ }\href {https://doi.org/10.1007/jhep04(2016)001}
  {\bibfield  {journal} {\bibinfo  {journal} {Journal of High Energy Physics}\
  }\textbf {\bibinfo {volume} {2016}},\ \bibinfo {pages} {1–25} (\bibinfo
  {year} {2016})}\BibitemShut {NoStop}%
\bibitem [{\citenamefont {Micklitz}\ \emph {et~al.}(2019)\citenamefont
  {Micklitz}, \citenamefont {Monteiro},\ and\ \citenamefont
  {Altland}}]{micklitz2019nonergodic}%
  \BibitemOpen
  \bibfield  {author} {\bibinfo {author} {\bibfnamefont {T.}~\bibnamefont
  {Micklitz}}, \bibinfo {author} {\bibfnamefont {F.}~\bibnamefont {Monteiro}},\
  and\ \bibinfo {author} {\bibfnamefont {A.}~\bibnamefont {Altland}},\
  }\bibfield  {title} {\bibinfo {title} {Nonergodic extended states in the
  sachdev-ye-kitaev model},\ }\href
  {https://doi.org/10.1103/PhysRevLett.123.125701} {\bibfield  {journal}
  {\bibinfo  {journal} {Phys. Rev. Lett.}\ }\textbf {\bibinfo {volume} {123}},\
  \bibinfo {pages} {125701} (\bibinfo {year} {2019})}\BibitemShut {NoStop}%
\bibitem [{\citenamefont {Kolodrubetz}\ \emph {et~al.}(2017)\citenamefont
  {Kolodrubetz}, \citenamefont {Sels}, \citenamefont {Mehta},\ and\
  \citenamefont {Polkovnikov}}]{kolodrubetz2017geometry}%
  \BibitemOpen
  \bibfield  {author} {\bibinfo {author} {\bibfnamefont {M.}~\bibnamefont
  {Kolodrubetz}}, \bibinfo {author} {\bibfnamefont {D.}~\bibnamefont {Sels}},
  \bibinfo {author} {\bibfnamefont {P.}~\bibnamefont {Mehta}},\ and\ \bibinfo
  {author} {\bibfnamefont {A.}~\bibnamefont {Polkovnikov}},\ }\bibfield
  {title} {\bibinfo {title} {Geometry and non-adiabatic response in quantum and
  classical systems},\ }\href
  {https://doi.org/https://doi.org/10.1016/j.physrep.2017.07.001} {\bibfield
  {journal} {\bibinfo  {journal} {Physics Reports}\ }\textbf {\bibinfo {volume}
  {697}},\ \bibinfo {pages} {1} (\bibinfo {year} {2017})}\BibitemShut {NoStop}%
\bibitem [{\citenamefont {Hunter-Jones}\ \emph {et~al.}(2018)\citenamefont
  {Hunter-Jones}, \citenamefont {Liu},\ and\ \citenamefont
  {Zhou}}]{hunter-jones2018on}%
  \BibitemOpen
  \bibfield  {author} {\bibinfo {author} {\bibfnamefont {N.}~\bibnamefont
  {Hunter-Jones}}, \bibinfo {author} {\bibfnamefont {J.}~\bibnamefont {Liu}},\
  and\ \bibinfo {author} {\bibfnamefont {Y.}~\bibnamefont {Zhou}},\ }\bibfield
  {title} {\bibinfo {title} {On thermalization in the {SYK} and supersymmetric
  {SYK} models},\ }\href {https://doi.org/10.1007%2Fjhep02%282018%29142}
  {\bibfield  {journal} {\bibinfo  {journal} {Journal of High Energy Physics}\
  }\textbf {\bibinfo {volume} {2018}} (\bibinfo {year} {2018})}\BibitemShut
  {NoStop}%
\bibitem [{\citenamefont {LeBlond}\ \emph {et~al.}(2021)\citenamefont
  {LeBlond}, \citenamefont {Sels}, \citenamefont {Polkovnikov},\ and\
  \citenamefont {Rigol}}]{leblond2021universality}%
  \BibitemOpen
  \bibfield  {author} {\bibinfo {author} {\bibfnamefont {T.}~\bibnamefont
  {LeBlond}}, \bibinfo {author} {\bibfnamefont {D.}~\bibnamefont {Sels}},
  \bibinfo {author} {\bibfnamefont {A.}~\bibnamefont {Polkovnikov}},\ and\
  \bibinfo {author} {\bibfnamefont {M.}~\bibnamefont {Rigol}},\ }\bibfield
  {title} {\bibinfo {title} {Universality in the onset of quantum chaos in
  many-body systems},\ }\href {https://doi.org/10.1103/PhysRevB.104.L201117}
  {\bibfield  {journal} {\bibinfo  {journal} {Phys. Rev. B}\ }\textbf {\bibinfo
  {volume} {104}},\ \bibinfo {pages} {L201117} (\bibinfo {year}
  {2021})}\BibitemShut {NoStop}%
\bibitem [{\citenamefont {{Cotler}}\ \emph {et~al.}(2017)\citenamefont
  {{Cotler}}, \citenamefont {{Gur-Ari}}, \citenamefont {{Hanada}},
  \citenamefont {{Polchinski}}, \citenamefont {{Saad}}, \citenamefont
  {{Shenker}}, \citenamefont {{Stanford}}, \citenamefont {{Streicher}},\ and\
  \citenamefont {{Tezuka}}}]{cotler2017black}%
  \BibitemOpen
  \bibfield  {author} {\bibinfo {author} {\bibfnamefont {J.~S.}\ \bibnamefont
  {{Cotler}}}, \bibinfo {author} {\bibfnamefont {G.}~\bibnamefont {{Gur-Ari}}},
  \bibinfo {author} {\bibfnamefont {M.}~\bibnamefont {{Hanada}}}, \bibinfo
  {author} {\bibfnamefont {J.}~\bibnamefont {{Polchinski}}}, \bibinfo {author}
  {\bibfnamefont {P.}~\bibnamefont {{Saad}}}, \bibinfo {author} {\bibfnamefont
  {S.~H.}\ \bibnamefont {{Shenker}}}, \bibinfo {author} {\bibfnamefont
  {D.}~\bibnamefont {{Stanford}}}, \bibinfo {author} {\bibfnamefont
  {A.}~\bibnamefont {{Streicher}}},\ and\ \bibinfo {author} {\bibfnamefont
  {M.}~\bibnamefont {{Tezuka}}},\ }\bibfield  {title} {\bibinfo {title} {{Black
  holes and random matrices}},\ }\href
  {https://doi.org/10.1007/JHEP05(2017)118} {\bibfield  {journal} {\bibinfo
  {journal} {Journal of High Energy Physics}\ }\textbf {\bibinfo {volume}
  {2017}},\ \bibinfo {eid} {118} (\bibinfo {year} {2017})}\BibitemShut
  {NoStop}%
\bibitem [{\citenamefont {Garc\'{\i}a-Garc\'{\i}a}\ \emph
  {et~al.}(2018{\natexlab{b}})\citenamefont {Garc\'{\i}a-Garc\'{\i}a},
  \citenamefont {Jia},\ and\ \citenamefont
  {Verbaarschot}}]{garcia-garcia2018universality}%
  \BibitemOpen
  \bibfield  {author} {\bibinfo {author} {\bibfnamefont {A.~M.}\ \bibnamefont
  {Garc\'{\i}a-Garc\'{\i}a}}, \bibinfo {author} {\bibfnamefont
  {Y.}~\bibnamefont {Jia}},\ and\ \bibinfo {author} {\bibfnamefont {J.~J.~M.}\
  \bibnamefont {Verbaarschot}},\ }\bibfield  {title} {\bibinfo {title}
  {Universality and thouless energy in the supersymmetric sachdev-ye-kitaev
  model},\ }\href {https://doi.org/10.1103/PhysRevD.97.106003} {\bibfield
  {journal} {\bibinfo  {journal} {Phys. Rev. D}\ }\textbf {\bibinfo {volume}
  {97}},\ \bibinfo {pages} {106003} (\bibinfo {year}
  {2018}{\natexlab{b}})}\BibitemShut {NoStop}%
\bibitem [{\citenamefont {{Gharibyan}}\ \emph {et~al.}(2018)\citenamefont
  {{Gharibyan}}, \citenamefont {{Hanada}}, \citenamefont {{Shenker}},\ and\
  \citenamefont {{Tezuka}}}]{gharibyan2018onset}%
  \BibitemOpen
  \bibfield  {author} {\bibinfo {author} {\bibfnamefont {H.}~\bibnamefont
  {{Gharibyan}}}, \bibinfo {author} {\bibfnamefont {M.}~\bibnamefont
  {{Hanada}}}, \bibinfo {author} {\bibfnamefont {S.~H.}\ \bibnamefont
  {{Shenker}}},\ and\ \bibinfo {author} {\bibfnamefont {M.}~\bibnamefont
  {{Tezuka}}},\ }\bibfield  {title} {\bibinfo {title} {{Onset of random matrix
  behavior in scrambling systems}},\ }\href
  {https://doi.org/10.1007/JHEP07(2018)124} {\bibfield  {journal} {\bibinfo
  {journal} {Journal of High Energy Physics}\ }\textbf {\bibinfo {volume}
  {2018}},\ \bibinfo {eid} {124} (\bibinfo {year} {2018})}\BibitemShut
  {NoStop}%
\bibitem [{Note1()}]{Note1}%
  \BibitemOpen
  \bibinfo {note} {Since the SFF is a spectral observable, \protect \emph
  {i.e.} it is based on eigenvalues, the relevant separation between regime
  \(III\) and regime \(IV\) is the one described by analytical predictions
  (vertical solid lines in Fig.~\ref {fig:AGP_plot_dillip_data}).}\BibitemShut
  {Stop}%
\bibitem [{\citenamefont {Winer}\ \emph {et~al.}(2020)\citenamefont {Winer},
  \citenamefont {Jian},\ and\ \citenamefont {Swingle}}]{winer2020exponential}%
  \BibitemOpen
  \bibfield  {author} {\bibinfo {author} {\bibfnamefont {M.}~\bibnamefont
  {Winer}}, \bibinfo {author} {\bibfnamefont {S.-K.}\ \bibnamefont {Jian}},\
  and\ \bibinfo {author} {\bibfnamefont {B.}~\bibnamefont {Swingle}},\
  }\bibfield  {title} {\bibinfo {title} {Exponential ramp in the quadratic
  sachdev-ye-kitaev model},\ }\href
  {https://doi.org/10.1103/PhysRevLett.125.250602} {\bibfield  {journal}
  {\bibinfo  {journal} {Phys. Rev. Lett.}\ }\textbf {\bibinfo {volume} {125}},\
  \bibinfo {pages} {250602} (\bibinfo {year} {2020})}\BibitemShut {NoStop}%
\bibitem [{\citenamefont {Liao}\ \emph {et~al.}(2020)\citenamefont {Liao},
  \citenamefont {Vikram},\ and\ \citenamefont {Galitski}}]{liao2020many}%
  \BibitemOpen
  \bibfield  {author} {\bibinfo {author} {\bibfnamefont {Y.}~\bibnamefont
  {Liao}}, \bibinfo {author} {\bibfnamefont {A.}~\bibnamefont {Vikram}},\ and\
  \bibinfo {author} {\bibfnamefont {V.}~\bibnamefont {Galitski}},\ }\bibfield
  {title} {\bibinfo {title} {Many-body level statistics of single-particle
  quantum chaos},\ }\href {https://doi.org/10.1103/PhysRevLett.125.250601}
  {\bibfield  {journal} {\bibinfo  {journal} {Phys. Rev. Lett.}\ }\textbf
  {\bibinfo {volume} {125}},\ \bibinfo {pages} {250601} (\bibinfo {year}
  {2020})}\BibitemShut {NoStop}%
\bibitem [{\citenamefont {Liao}\ and\ \citenamefont
  {Galitski}(2022)}]{liao2022universal}%
  \BibitemOpen
  \bibfield  {author} {\bibinfo {author} {\bibfnamefont {Y.}~\bibnamefont
  {Liao}}\ and\ \bibinfo {author} {\bibfnamefont {V.}~\bibnamefont
  {Galitski}},\ }\bibfield  {title} {\bibinfo {title} {Universal dephasing
  mechanism of many-body quantum chaos},\ }\href
  {https://doi.org/10.1103/PhysRevResearch.4.L012037} {\bibfield  {journal}
  {\bibinfo  {journal} {Phys. Rev. Research}\ }\textbf {\bibinfo {volume}
  {4}},\ \bibinfo {pages} {L012037} (\bibinfo {year} {2022})}\BibitemShut
  {NoStop}%
\bibitem [{\citenamefont {Chan}\ \emph {et~al.}(2018)\citenamefont {Chan},
  \citenamefont {De~Luca},\ and\ \citenamefont {Chalker}}]{chan2018spectral}%
  \BibitemOpen
  \bibfield  {author} {\bibinfo {author} {\bibfnamefont {A.}~\bibnamefont
  {Chan}}, \bibinfo {author} {\bibfnamefont {A.}~\bibnamefont {De~Luca}},\ and\
  \bibinfo {author} {\bibfnamefont {J.~T.}\ \bibnamefont {Chalker}},\
  }\bibfield  {title} {\bibinfo {title} {Spectral statistics in spatially
  extended chaotic quantum many-body systems},\ }\href
  {https://doi.org/10.1103/PhysRevLett.121.060601} {\bibfield  {journal}
  {\bibinfo  {journal} {Phys. Rev. Lett.}\ }\textbf {\bibinfo {volume} {121}},\
  \bibinfo {pages} {060601} (\bibinfo {year} {2018})}\BibitemShut {NoStop}%
\bibitem [{Note2()}]{Note2}%
  \BibitemOpen
  \bibinfo {note} {The case \(N = 22\) seems to be smallest value for which a
  scaling behavior is observed~\cite {monteiro2021minimal}.}\BibitemShut
  {Stop}%
\bibitem [{\citenamefont {Sierant}\ and\ \citenamefont
  {Zakrzewski}(2021)}]{sierant2021challenges}%
  \BibitemOpen
  \bibfield  {author} {\bibinfo {author} {\bibfnamefont {P.}~\bibnamefont
  {Sierant}}\ and\ \bibinfo {author} {\bibfnamefont {J.}~\bibnamefont
  {Zakrzewski}},\ }\href {https://doi.org/10.48550/ARXIV.2109.13608} {\bibinfo
  {title} {Challenges to observation of many-body localization}} (\bibinfo
  {year} {2021})\BibitemShut {NoStop}%
\bibitem [{\citenamefont {Tang}\ and\ \citenamefont
  {Khaymovich}(2021)}]{tang2021non-ergodic}%
  \BibitemOpen
  \bibfield  {author} {\bibinfo {author} {\bibfnamefont {W.}~\bibnamefont
  {Tang}}\ and\ \bibinfo {author} {\bibfnamefont {I.~M.}\ \bibnamefont
  {Khaymovich}},\ }\href {https://doi.org/10.48550/ARXIV.2112.09700} {\bibinfo
  {title} {Non-ergodic delocalized phase with poisson level statistics}}
  (\bibinfo {year} {2021})\BibitemShut {NoStop}%
\bibitem [{\citenamefont {Atas}\ \emph
  {et~al.}(2013{\natexlab{a}})\citenamefont {Atas}, \citenamefont {Bogomolny},
  \citenamefont {Giraud},\ and\ \citenamefont {Roux}}]{Atas2013}%
  \BibitemOpen
  \bibfield  {author} {\bibinfo {author} {\bibfnamefont {Y.~Y.}\ \bibnamefont
  {Atas}}, \bibinfo {author} {\bibfnamefont {E.}~\bibnamefont {Bogomolny}},
  \bibinfo {author} {\bibfnamefont {O.}~\bibnamefont {Giraud}},\ and\ \bibinfo
  {author} {\bibfnamefont {G.}~\bibnamefont {Roux}},\ }\bibfield  {title}
  {\bibinfo {title} {Distribution of the ratio of consecutive level spacings in
  random matrix ensembles},\ }\href
  {https://doi.org/10.1103/PhysRevLett.110.084101} {\bibfield  {journal}
  {\bibinfo  {journal} {Phys. Rev. Lett.}\ }\textbf {\bibinfo {volume} {110}},\
  \bibinfo {pages} {084101} (\bibinfo {year} {2013}{\natexlab{a}})}\BibitemShut
  {NoStop}%
\bibitem [{\citenamefont {Atas}\ \emph
  {et~al.}(2013{\natexlab{b}})\citenamefont {Atas}, \citenamefont {Bogomolny},
  \citenamefont {Giraud}, \citenamefont {Vivo},\ and\ \citenamefont
  {Vivo}}]{Atas2013a}%
  \BibitemOpen
  \bibfield  {author} {\bibinfo {author} {\bibfnamefont {Y.}~\bibnamefont
  {Atas}}, \bibinfo {author} {\bibfnamefont {E.}~\bibnamefont {Bogomolny}},
  \bibinfo {author} {\bibfnamefont {O.}~\bibnamefont {Giraud}}, \bibinfo
  {author} {\bibfnamefont {P.}~\bibnamefont {Vivo}},\ and\ \bibinfo {author}
  {\bibfnamefont {E.}~\bibnamefont {Vivo}},\ }\bibfield  {title} {\bibinfo
  {title} {Joint probability densities of level spacing ratios in random
  matrices},\ }\href@noop {} {\bibfield  {journal} {\bibinfo  {journal} {J.
  Phys. A}\ }\textbf {\bibinfo {volume} {46}},\ \bibinfo {pages} {355204}
  (\bibinfo {year} {2013}{\natexlab{b}})}\BibitemShut {NoStop}%
\bibitem [{\citenamefont {Rela\~no}\ \emph {et~al.}(2002)\citenamefont
  {Rela\~no}, \citenamefont {G\'omez}, \citenamefont {Molina}, \citenamefont
  {Retamosa},\ and\ \citenamefont {Faleiro}}]{Relano2002}%
  \BibitemOpen
  \bibfield  {author} {\bibinfo {author} {\bibfnamefont {A.}~\bibnamefont
  {Rela\~no}}, \bibinfo {author} {\bibfnamefont {J.~M.~G.}\ \bibnamefont
  {G\'omez}}, \bibinfo {author} {\bibfnamefont {R.~A.}\ \bibnamefont {Molina}},
  \bibinfo {author} {\bibfnamefont {J.}~\bibnamefont {Retamosa}},\ and\
  \bibinfo {author} {\bibfnamefont {E.}~\bibnamefont {Faleiro}},\ }\bibfield
  {title} {\bibinfo {title} {Quantum chaos and $1/f$ noise},\ }\href
  {https://doi.org/10.1103/PhysRevLett.89.244102} {\bibfield  {journal}
  {\bibinfo  {journal} {Phys. Rev. Lett.}\ }\textbf {\bibinfo {volume} {89}},\
  \bibinfo {pages} {244102} (\bibinfo {year} {2002})}\BibitemShut {NoStop}%
\bibitem [{\citenamefont {Faleiro}\ \emph {et~al.}(2004)\citenamefont
  {Faleiro}, \citenamefont {G\'omez}, \citenamefont {Molina}, \citenamefont
  {Mu\~noz}, \citenamefont {Rela\~no},\ and\ \citenamefont
  {Retamosa}}]{Faleiro2004}%
  \BibitemOpen
  \bibfield  {author} {\bibinfo {author} {\bibfnamefont {E.}~\bibnamefont
  {Faleiro}}, \bibinfo {author} {\bibfnamefont {J.~M.~G.}\ \bibnamefont
  {G\'omez}}, \bibinfo {author} {\bibfnamefont {R.~A.}\ \bibnamefont {Molina}},
  \bibinfo {author} {\bibfnamefont {L.}~\bibnamefont {Mu\~noz}}, \bibinfo
  {author} {\bibfnamefont {A.}~\bibnamefont {Rela\~no}},\ and\ \bibinfo
  {author} {\bibfnamefont {J.}~\bibnamefont {Retamosa}},\ }\bibfield  {title}
  {\bibinfo {title} {Theoretical derivation of $1/f$ noise in quantum chaos},\
  }\href {https://doi.org/10.1103/PhysRevLett.93.244101} {\bibfield  {journal}
  {\bibinfo  {journal} {Phys. Rev. Lett.}\ }\textbf {\bibinfo {volume} {93}},\
  \bibinfo {pages} {244101} (\bibinfo {year} {2004})}\BibitemShut {NoStop}%
\bibitem [{\citenamefont {G\'omez}\ \emph {et~al.}(2005)\citenamefont
  {G\'omez}, \citenamefont {Rela\~no}, \citenamefont {Retamosa}, \citenamefont
  {Faleiro}, \citenamefont {Salasnich}, \citenamefont
  {Vrani\ifmmode~\check{c}\else \v{c}\fi{}ar},\ and\ \citenamefont
  {Robnik}}]{Gomez2005}%
  \BibitemOpen
  \bibfield  {author} {\bibinfo {author} {\bibfnamefont {J.~M.~G.}\
  \bibnamefont {G\'omez}}, \bibinfo {author} {\bibfnamefont {A.}~\bibnamefont
  {Rela\~no}}, \bibinfo {author} {\bibfnamefont {J.}~\bibnamefont {Retamosa}},
  \bibinfo {author} {\bibfnamefont {E.}~\bibnamefont {Faleiro}}, \bibinfo
  {author} {\bibfnamefont {L.}~\bibnamefont {Salasnich}}, \bibinfo {author}
  {\bibfnamefont {M.}~\bibnamefont {Vrani\ifmmode~\check{c}\else
  \v{c}\fi{}ar}},\ and\ \bibinfo {author} {\bibfnamefont {M.}~\bibnamefont
  {Robnik}},\ }\bibfield  {title} {\bibinfo {title}
  {$1/{f}^{\ensuremath{\alpha}}$ noise in spectral fluctuations of quantum
  systems},\ }\href {https://doi.org/10.1103/PhysRevLett.94.084101} {\bibfield
  {journal} {\bibinfo  {journal} {Phys. Rev. Lett.}\ }\textbf {\bibinfo
  {volume} {94}},\ \bibinfo {pages} {084101} (\bibinfo {year}
  {2005})}\BibitemShut {NoStop}%
\bibitem [{\citenamefont {Salasnich}(2005)}]{Salasnich2005}%
  \BibitemOpen
  \bibfield  {author} {\bibinfo {author} {\bibfnamefont {L.}~\bibnamefont
  {Salasnich}},\ }\bibfield  {title} {\bibinfo {title} {Colored noise in
  quantum chaos},\ }\href {https://doi.org/10.1103/PhysRevE.71.047202}
  {\bibfield  {journal} {\bibinfo  {journal} {Phys. Rev. E}\ }\textbf {\bibinfo
  {volume} {71}},\ \bibinfo {pages} {047202} (\bibinfo {year}
  {2005})}\BibitemShut {NoStop}%
\bibitem [{\citenamefont {Santhanam}\ and\ \citenamefont
  {Bandyopadhyay}(2005)}]{Santhanam2005}%
  \BibitemOpen
  \bibfield  {author} {\bibinfo {author} {\bibfnamefont {M.~S.}\ \bibnamefont
  {Santhanam}}\ and\ \bibinfo {author} {\bibfnamefont {J.~N.}\ \bibnamefont
  {Bandyopadhyay}},\ }\bibfield  {title} {\bibinfo {title} {Spectral
  fluctuations and $1/f$ noise in the order-chaos transition regime},\ }\href
  {https://doi.org/10.1103/PhysRevLett.95.114101} {\bibfield  {journal}
  {\bibinfo  {journal} {Phys. Rev. Lett.}\ }\textbf {\bibinfo {volume} {95}},\
  \bibinfo {pages} {114101} (\bibinfo {year} {2005})}\BibitemShut {NoStop}%
\bibitem [{\citenamefont {Rela\~no}(2008)}]{Relano2008}%
  \BibitemOpen
  \bibfield  {author} {\bibinfo {author} {\bibfnamefont {A.}~\bibnamefont
  {Rela\~no}},\ }\bibfield  {title} {\bibinfo {title} {Chaos-assisted tunneling
  and $1/{f}^{\ensuremath{\alpha}}$ spectral fluctuations in the order-chaos
  transition},\ }\href {https://doi.org/10.1103/PhysRevLett.100.224101}
  {\bibfield  {journal} {\bibinfo  {journal} {Phys. Rev. Lett.}\ }\textbf
  {\bibinfo {volume} {100}},\ \bibinfo {pages} {224101} (\bibinfo {year}
  {2008})}\BibitemShut {NoStop}%
\bibitem [{\citenamefont {Faleiro}\ \emph {et~al.}(2006)\citenamefont
  {Faleiro}, \citenamefont {Kuhl}, \citenamefont {Molina}, \citenamefont
  {Muñoz}, \citenamefont {Relaño},\ and\ \citenamefont
  {Retamosa}}]{Faleiro2006}%
  \BibitemOpen
  \bibfield  {author} {\bibinfo {author} {\bibfnamefont {E.}~\bibnamefont
  {Faleiro}}, \bibinfo {author} {\bibfnamefont {U.}~\bibnamefont {Kuhl}},
  \bibinfo {author} {\bibfnamefont {R.}~\bibnamefont {Molina}}, \bibinfo
  {author} {\bibfnamefont {L.}~\bibnamefont {Muñoz}}, \bibinfo {author}
  {\bibfnamefont {A.}~\bibnamefont {Relaño}},\ and\ \bibinfo {author}
  {\bibfnamefont {J.}~\bibnamefont {Retamosa}},\ }\bibfield  {title} {\bibinfo
  {title} {Power spectrum analysis of experimental sinai quantum billiards},\
  }\href {https://doi.org/https://doi.org/10.1016/j.physleta.2006.05.029}
  {\bibfield  {journal} {\bibinfo  {journal} {Phys. Lett. A}\ }\textbf
  {\bibinfo {volume} {358}},\ \bibinfo {pages} {251 } (\bibinfo {year}
  {2006})}\BibitemShut {NoStop}%
\bibitem [{\citenamefont {Mur-Petit}\ and\ \citenamefont
  {Molina}(2015)}]{Mur2015}%
  \BibitemOpen
  \bibfield  {author} {\bibinfo {author} {\bibfnamefont {J.}~\bibnamefont
  {Mur-Petit}}\ and\ \bibinfo {author} {\bibfnamefont {R.~A.}\ \bibnamefont
  {Molina}},\ }\bibfield  {title} {\bibinfo {title} {Spectral statistics of
  molecular resonances in erbium isotopes: How chaotic are they?},\ }\href
  {https://doi.org/10.1103/PhysRevE.92.042906} {\bibfield  {journal} {\bibinfo
  {journal} {Phys. Rev. E}\ }\textbf {\bibinfo {volume} {92}},\ \bibinfo
  {pages} {042906} (\bibinfo {year} {2015})}\BibitemShut {NoStop}%
\end{thebibliography}
\end{document}